\documentclass[prx,eqsecnum,twocolumn,aps,
showpacs,amsmath,amssymb,floatfix,
longbibliography,superscriptaddress]{revtex4}
\usepackage{graphicx}
\DeclareGraphicsExtensions{.pdf}
\usepackage{amsmath,amssymb,bbold,bm,color}
\usepackage{float}
\usepackage{epstopdf}
\usepackage[dvipsnames]{xcolor}

\newcommand{\bk}{{\bm k}}

\newcommand{\bp}{{\bm p}}
\newcommand{\br}{{\bm r}}

\newcommand{\bB}{{\bm B}}

\newcommand{\bA}{{\bm A}}

\newcommand{\bsig}{{\bm \sigma}}
\newcommand{\btau}{{\bm \tau}}

\newcommand{\cT}{{\cal T}}
\newcommand{\cK}{{\cal K}}

\newcommand{\cH}{{\cal H}}

\newcommand{\bee}{\begin{equation}}
\newcommand{\ee}{\end{equation}}

\def\sgn{\mathop{\rm sgn}\nolimits}

\begin{document}

\title{Black hole on a chip: proposal for a physical realization of
  the SYK model in a solid-state system}

\author{D.I. Pikulin}
\affiliation{Station Q, Microsoft Research, Santa Barbara, California 93106-6105, USA}
\author{M. Franz}
\affiliation{Department of Physics and Astronomy, University of
British Columbia, Vancouver, BC, Canada V6T 1Z1}
\affiliation{Quantum Matter Institute, University of British Columbia, Vancouver BC, Canada V6T 1Z4}

\date{\today}

\begin{abstract} 
System of Majorana zero modes with random infinite range interactions --
the Sachdev-Ye-Kitaev (SYK) model -- is thought to exhibit an
intriguing relation to the horizons of
extremal black holes in two-dimensional anti-de Sitter (AdS$_2$)
space. This connection provides a rare example of holographic duality
between a solvable quantum-mechanical model and dilaton gravity. Here
we propose a physical realization of the SYK model in a solid state
system. The proposed setup employs the Fu-Kane superconductor realized
at the interface  between a three dimensional topological insulator (TI)
and an ordinary superconductor. The requisite $N$ Majorana zero modes are
bound to a nanoscale hole fabricated  in the superconductor that is
threaded by $N$ quanta of magnetic flux. We show that when the system
is tuned to the surface neutrality point (i.e. chemical potential coincident
with the Dirac point of the TI surface state) and the hole has
sufficiently irregular shape, the Majorana zero modes are described by
the SYK Hamiltonian. We perform extensive numerical simulations to
demonstrate that the system indeed exhibits physical properties
expected of the SYK model, including thermodynamic quantities and
two-point as well as four-point correlators, and discuss ways in which
these can be observed experimentally.  

\end{abstract}

\date{\today}

\maketitle

\section{Introduction}
Models of particles with infinite-range interactions have a long
history in nuclear physics dating back to the pioneering works of  Wigner and Dyson
\cite{Wigner1951,Dyson1962}  and in condensed matter
physics in studies describing spin glass and spin
liquid states of matter \cite{SK1975,SY1996,Camjayi2003} . More
recently, Kitaev \cite{Kitaev2015,Maldacena2016} formulated and studied a  Majorana
fermion version of the model  with all-to-all
random  interactions  first proposed by Sachdev and Ye
\cite{SY1996}. The resulting SYK model, defined by the Hamiltonian
(\ref{hsyk}) below, is solvable in the limit of large number $N$ of
fermions  and exhibits a host of intriguing properties. The SYK model
is believed to be holographic dual of extremal black hole horizons
in  AdS$_2$ and has been argued to possess remarkable connections to
information theory, many-body thermalization and quantum chaos
\cite{Sachdev2015,Maldacena2016b,Hosur2016,Polchinski2016,Verbaar2016,Xu2016}.
Various extensions of the SYK model have been put forth  containing
supersymmetry \cite{Fu2016}, interesting quantum phase transition
\cite{Altman2016,Bi2017},  higher dimensional extensions
\cite{Gu2016,Berkooz2016}, as well as a version that does not require
randomness \cite{Witten2016}. Given its fascinating properties it would
be of obvious interest to have an experimental realization of the SYK
model or its variants. Thus far a realization of the Sachdev-Ye model
(with complex fermions) has been proposed using ultracold gases
\cite{Danshita2016} and a protocol for digital quantum simulation of
both the complex and Majorana fermion versions of the model has been
discussed \cite{Garcia2016}. A natural realization of the SYK model in
a solid state system is thus far lacking.   

Recent years have witnesed numerous proposals for experimental realizations of unpaired
Majorana zero modes in solid state systems
\cite{Alicea2012,Beenakker2012,Leijnse2012,Stanescu2013,Elliott2015},  with compelling
experimental evidence for their existence gradually mounting in several
distinct platforms \cite{Mourik2012,Das2012,Deng2012,Rokhinson2012,Finck2013,Hart2014,Nadj-Perge2014,Jia2016a,Jia2016b}. The purpose of this paper is to propose a
physical realization of the SYK model in one of these platforms. The SYK
Hamiltonian we wish to implement is given by  
\begin{equation}\label{hsyk}
{\cal H}_{\rm SYK}=\sum_{i<j<k<l} J_{ijkl}\chi_i\chi_j \chi_k\chi_l,
\end{equation}
where $J_{ijkl}$ are random independent coupling constants and
$\chi_j$ represent the Majorana zero mode operators that obey
the canonical anticommutation relations
\begin{equation}\label{hsyk2}
\{\chi_i,\chi_j\}=\delta_{ij}, \ \ \  \chi_j^\dagger=\chi_j.
\end{equation}
\begin{figure}[t]
\includegraphics[width = 7.5cm]{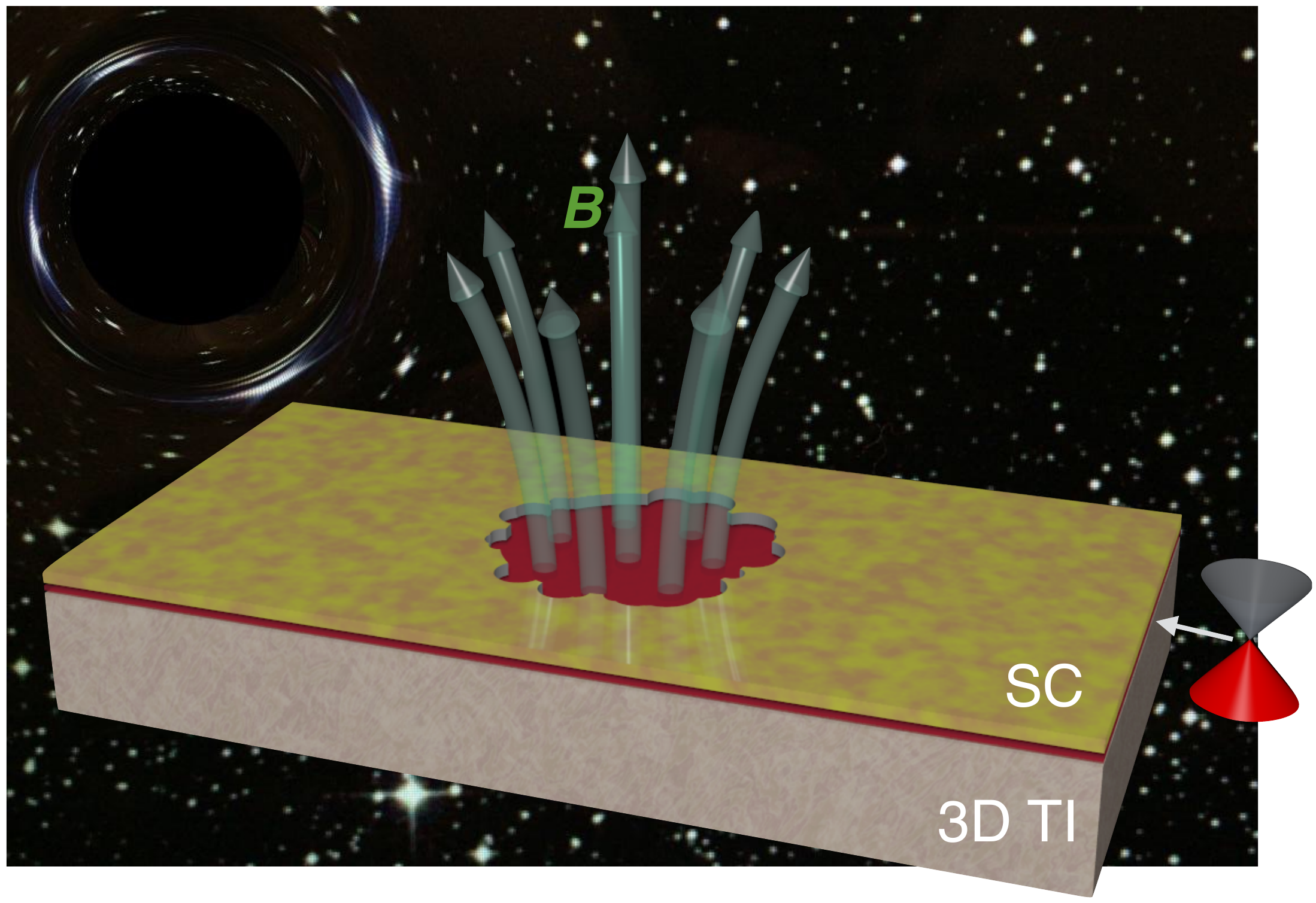}
\caption{The proposed setup for a solid-state realization of the SYK
  model. 
}\label{fig1}
\end{figure}

The proposed device,  depicted in Fig.\
\ref{fig1}, employs an interface between a 3D TI and an ordinary
superconductor such as Nb or Pb. Fu and Kane \cite{FuKane2008} showed theoretically that magnetic vortices in such an
interface host unpaired Majorana zero modes and signatures consistent
with this prediction have been reported in Bi$_2$Te$_3$/NbSe$_2$
heterostructures \cite{Jia2016a,Jia2016b}. Under ordinary circumstances these vortices tend to form an
Abrikosov lattice  and the low-energy effective theory is dominated by
two-fermion terms $iK_{ij}\chi_i\chi_j$ with the hoping amplitudes
$K_{ij}$ decaying exponentially with the distance between vortex sites
$|\br_i-\br_j|$. Four-fermion interaction terms of the type required
to implement the SYK Hamiltonian (\ref{hsyk}) are generically  also present but
are subdominant and also decay exponentially with distance. Realizing the SYK
model in this setup therefore entails two key
challenges: (i) one must find a way to suppress the two-fermion
tunneling terms and (ii) render the four-fermion interactions
effectively infinite-ranged. In addition the four-fermion coupling
constants $J_{ijkl}$  must be sufficiently random.  
In the following we show how these challenges can be overcome by
judicious engineering of various aspects of the device depicted in
Fig.\ \ref{fig1}. 

The first challenge can
be met by tuning the surface state of the TI into its global neutrality point
such that the chemical potential $\mu$ lies at the Dirac point. At the
neutrality point the interface superconductor is known to acquire an
extra chiral symmetry  which prohibits any two-fermion terms
\cite{TeoKane2010}. In other words, the symmetry requires  $K_{ij}=0$ and
the low-energy Hamiltonian is then dominated by the four-fermion terms \cite{Chiu2015}.
The second requirement of effectively infinite-ranged interactions can
be satisfied
by localizing all Majorana zero modes in the same region of
space. In our proposed device this is achieved by fabricating a hole in the SC layer
as illustrated in Fig.  \ref{fig1}. If the sample is cooled in a weak
applied 
magnetic field an integer number  $N$ of magnetic flux quanta can be 
trapped in the hole. The SC phase $\theta$ will then wind by $2\pi N$
around the hole, forming effectively an $N$-fold vortex with $N$
Majorana zero modes bound to the hole.  If furthermore the hole is
designed to have an irregular shape  the Majorana wavefunctions will
have random spatial structure and their overlaps will give rise to the required randomness
in the coupling constants $J_{ijkl}$. This randomness
  is related to random classical trajectories inside such a hole, or
  `billiards' as it is commonly called in the quantum chaos literature
  \cite{gutzwiller1990chaos, haake2013quantum}. We note a related
  proposal to realize the SYK model using semiconductor quantum wires
  coupled to a disordered quantum dot advanced in the recent work \cite{Alicea2017}. 

In the rest of the paper we provide the necessary background on our
proposed system and support its relation to the SYK model by physical
arguments and by detailed model calculations.
We first review the Fu-Kane model \cite{FuKane2008} for the TI/SC interface
and numerically calculate the Majorana wavefunctions localized in a
hole threaded by $N$ magnetic flux quanta in the presence of
disorder. Assuming that the constituent electrons interact via
screened Coulomb potential we then explicitly calculate the
four-fermion coupling constants  $J_{ijkl}$ between the Majorana zero
modes. We finally use these as input data for the many-body  Majorana
Hamiltonian which we diagonalize numerically for $N$ up to 32 and
study its thermodynamic properties, level statistics, as well as  two- and four-point
correlators. We show that these behave precisely as expected of the
SYK model with random independent couplings. We also discuss the
effect of small residual two-fermion terms that will inevitably be
present in a realistic device and propose ways to experimentally
detect signatures of the SYK physics using tunneling spectroscopy.


\section{SYK model from Interacting Majorana zero modes at the TI/SC interface}

\subsection{The Fu-Kane superconductor}

The surface of a canonical 3D TI, such as  Bi$_2$Se$_3$, hosts a single massless  Dirac
fermion protected by time reversal symmetry.  When placed in the proximity
of an ordinary superconductor the surface state is described by
the Fu-Kane Hamiltonian \cite{FuKane2008}
\begin{equation}\label{abel1}
{\cal H}_{\rm FK}=\int d^2r \hat{\Psi}^\dagger_\br H_{\rm FK}(\br)\hat{\Psi}_\br,
\end{equation}
where
$\hat\Psi_\br=(c_{\uparrow\br},c_{\downarrow\br},c^\dagger_{\downarrow\br},-c^\dagger_{\uparrow\br})^T$
is the Nambu spinor and 
\begin{equation}\label{sym1}
H_{\rm FK}= \tau^z\left[v_F\bsig\cdot\left(\bp-\tau^z{e\over c}\bA\right)-\mu\right]+\tau^x\Delta_1-\tau^y\Delta_2.
\end{equation}
Here $v_F$ is the velocity of the surface state, $\bp=-i\hbar\nabla$
denotes the momentum operator, $\Delta=\Delta_1+i\Delta_2$ is the
SC order parameter and $\bsig$, $\btau$ are Pauli matrices in spin and
Nambu spaces, respectively. To describe the geometry depicted in Fig.\
\ref{fig1} we take
\begin{equation}\label{gap1}
\Delta(\br)=\left\{
\begin{array}{ll}
0, & r<R(\varphi) \\
\Delta_0 e^{iN\varphi},  & r>R(\varphi),
\end{array}
\right.
\end{equation}
where $\varphi$ is the polar angle and $R(\varphi)$ denotes the hole
radius. The vector potential is taken to yield total flux through the
hole  $\oint_C d{\bf l}\cdot\bA=N\Phi_0$ with $\Phi_0=hc/2e$ the SC
flux quantum and the contour $C$ taken to encircle the hole at a
radius well  beyond the effective magnetic penetration depth of the SC film
$\lambda_{\rm eff}=2\lambda_L^2/d$. (Here $\lambda_L$ is the London
penetration depth of the bulk SC and $d$ the film thickness.)

Hamiltonian (\ref{sym1}) respects the particle-hole symmetry generated by $\Xi=\sigma^y\tau^y \cK$ 
($\Xi^2=+1$), where $\cK$ denotes complex conjugation. For a purely real
gap function $\Delta$ and zero magnetic field $\bB=\nabla\times\bA$,
it also obeys the physical time reversal symmetry $\Theta=i\sigma^y \cK$
($\Theta^2=-1$). In the presence of vortices $\Delta$ becomes complex
and the time reversal symmetry is broken. The Fu-Kane model with vortices
therefore falls into symmetry class D in the Altland-Zirnbauer
classification \cite{Altland1997} which, in accordance with Ref.\ \cite{TeoKane2010}, implies a Z$_2$ classification for the zero modes associated with vortices. Physically, this means that a system with total vorticity $N_V$ will have $N=(N_V\! \mod 2)$ exact zero modes, in accord with the expectation that any even number of Majorana zero modes will generically hybridize and form  complex fermions at non-zero energy.

When $\mu=0$, Hamiltonian (\ref{sym1}) also respects a fictitious time
reversal symmetry generated by $\Sigma=\sigma^x\tau^x \cK$
($\Sigma^2=+1$). 
{
It is important to note that unlike the physical time reversal  this
symmetry remains valid even in the presence of the applied magnetic
field and vortices.} At the
neutrality point, the two symmetries $\Xi$ and $\Sigma$ define
a BDI class with chiral symmetry
$\Pi=\Xi\Sigma=\sigma^z\tau^z$. This, in accordance with  Ref.\
\cite{TeoKane2010}, implies an integer classification of zero modes
associated with point defects. A system with total vorticity $N_V$
will thus exhibit $N=N_V$ exact zero modes, irrespective of the precise
geometric arrangement of the individual vortices and other
details. This remains true in the presence of any disorder that does
not break  the $\Sigma$ symmetry. Specifically randomness in
$v_F$ and $\Delta$ will not split the zero modes but random
contributions to $\mu$ will. 

Another way to establish the existence of exact zero modes  in the
Hamiltonian (\ref{sym1}) with $\mu=0$ is to recognize it as a version
of the Jackiw-Rossi Hamiltonian \cite{Jackiw1981} well known in
particle physics. An index theorem for this
Hamiltonian, conjectured by Jackiw and Rossi  and later proven by
Weinberg  \cite{Weinberg1981}, equates the number $N$
of its exact zero modes in region ${\cal M}$  to the total vorticity
$N_V={1\over 2\pi}\oint_{\partial{\cal M}}d{\bf l}\cdot\nabla\theta$
contained in that region. A region threaded by $N_V$ magnetic flux
quanta will thus contain $N$ exact zero modes. 

{
In the geometry of Fig.\ \ref{fig1} the Majorana modes discussed above
can equivalently be viewed as living at the boundary between a
magnetically gapped TI surface on the inside and a SC region on the
outside of the hole. The existence of such modes is well known and has
been discussed in several papers \cite{Fu2009,Akhmerov2009}. }

The existence and properties of the zero modes in the Fu-Kane
Hamiltonian have been extensively tested by analytic and numerical
approaches for a single vortex \cite{FuKane2008}, pair of vortices
\cite{cheng2009,cheng2010},
periodic Abrikosov lattices \cite{Liu2015,Murray2015} as
well as  the ``giant vortex'' geometry \cite{pikulin2015} similar to our proposed
setup. This body of work firmly establishes the existence of exact Majorana
zero modes for $\mu=0$ in accordance with the Jackiw-Rossi-Weinberg
index theorem. Away from neutrality it is found that the zero modes
are split due to two-fermion tunneling terms $K_{ij}\propto\mu$ where
the constant of proportionality is related to the wavefunction overlap
between $\chi_i$ and $\chi_j$. In addition, it has been found that for
a singly quantized vortex at neutrality the zero mode is separated
from the rest of the spectrum by a gap $\sim\Delta_0$ where $\Delta_0$
is the SC gap magnitude far from the vortex. We shall see that for a
judiciously chosen hole size this convenient hierarchy of energy
scales remains in place with $N$ zero modes separated by a large gap
from the rest of the spectrum.

\subsection{Low-energy effective theory} 
Having established a convenient platform that hosts $N$ Majorana zero
modes with wavefunctions localized in the same region of space we now
proceed to derive the effective low-energy theory in terms of the
Majorana zero mode operators $\chi_j$. To this end we write the full
second-quantized Hamiltonian of the system as
\begin{equation}\label{h3}
{\cal H}={\cal H}_{\rm FK}^{(N)}+\delta{\cal H}_{\rm FK}+{\cal H}_{\rm int}.
\end{equation}
Here ${\cal H}_{\rm FK}^{(N)}$ stands for the part of the Fu-Kane
Hamiltonian (\ref{sym1}) that obeys the fictitious time-reversal
symmetry $\Sigma$ and exhibits therefore $N$ exact zero
modes. $\delta{\cal H}_{\rm FK}$ contains all the remaining
fermion bilinears that break $\Sigma$ such as the chemical
potential term.
${\cal H}_{\rm int}$ defines the four-fermion interactions that have
been ignored thus far but will play pivotal role in the physics of the
SYK model we wish to study. We assume that electrons are subject to
screened Coulomb interactions described by  
\begin{equation}\label{bb2}
{\cal H}_{\rm int}={1\over 2}\int\int d^2r d^2r'\hat{\rho}(\br)V(\br-\br')\hat{\rho}(\br'),
\end{equation}
where $V(\br)$ is the interaction potential and
$\hat{\rho}(\br)=c^\dagger_{\sigma\br}c_{\sigma\br}$ is the electron charge density
operator.  

Now imagine we have
solved the single-electron problem defined by Hamiltonian ${H}_{\rm
  FK}^{(N)}$ for the device geometry sketched in Fig.\ \ref{fig1} with
$N$ flux quanta threaded through the hole. We
thus have the complete set of single-particle eigenfunctions $\Phi_n(\br)$  and
eigenenergies $\varepsilon_n$ of $H_{\rm FK}^{(N)}$.  The corresponding second
quantized Hamiltonian can then be written in a diagonal form 
\begin{equation}\label{abel11}
{\cal H}_{\rm FK}^{(N)}={\sum_n}'
\varepsilon_n\hat{\psi}^\dagger_n\hat{\psi}_n +E_g
\end{equation}
 where 
\begin{equation}\label{abel12}
\hat{\psi}_n=\int d^2r \Phi_n^\dagger(\br)\hat{\Psi}_\br
\end{equation}
is the eigenmode operator belonging to the eigenvalue $\varepsilon_n$. The sum over $n$ is restricted to the
positive energy eigenvalues and $E_g$ is a constant representing the ground
state energy. At the neutrality point, according to our preceding
discussion,  $N$ of the $\hat{\psi}_n$ eigenmodes coincide with the
exact zero modes mandated by the Jackiw-Rossi-Weinberg index
theorem. We denote these $\chi_j$ with  $j=1\dots N$. Because
 $\varepsilon_j=0$ these modes do not contribute to the Hamiltonian
(\ref{abel11}). The zero mode eigenfunctions $\Phi_j(\br)$
can be chosen as eigenstates of the p-h symmetry generator
$\Xi$. They then satisfy the reality condition 
\begin{equation}\label{real1}
\sigma^y\tau^y\Phi_j ^*(\br)=\Phi_j(\br)
\end{equation}
 which implies that $\chi_j^\dagger=\chi_j$;
the zero modes are Majorana operators. 

As noted before the $N$ zero modes are separated by a gap from the
rest of the spectrum. As long as $\delta{\cal H}_{\rm FK}$ and ${\cal
  H}_{\rm int}$ remain small compared to this gap we may construct the
effective low-energy theory of the system by simply projecting
onto the part of the Hilbert space generated by $N$ Majorana zero
modes. In practical terms this is accomplished by inverting Eq.\
(\ref{abel12}) to obtain 
\begin{equation}\label{abel13}
\hat{\Psi}_\br=\sum_n\Phi_n(\br)\hat{\psi}_n, 
\end{equation}
then substituting $\hat{\Psi}_\br$ into $\delta{\cal H}_{\rm FK}$ and ${\cal
  H}_{\rm int}$ and retaining only those terms that contain zero mode
operators $\chi_j$ but no finite-energy eigenmodes. We thus obtain
\begin{equation}\label{heff}
{\cal H}_{\rm eff}={i\over 2!}\sum_{i,j}
\tilde{K}_{ij}\chi_i\chi_j+{1\over 4!}\sum_{i,j,k,l} \tilde{J}_{ijkl}\chi_i\chi_j \chi_k\chi_l,
\end{equation}
where 
\begin{eqnarray}\label{heffK}
i\tilde{K}_{ij}&=&2!\int d^2r \Phi_i^\dagger(\br)\delta H_{\rm FK}(\br)
           \Phi_j(\br), \\
\tilde{J}_{ijkl}&=&{4!\over 2}\int\int d^2r d^2r'
            \rho_{ij}(\br)V(\br-\br')\rho_{lk}(\br'), \ \ \ \ \  \label{heffJ}
\end{eqnarray}
and $\rho_{ij}(\br)={i\over 2}\Phi_i^\dagger(\br)\tau^z\Phi_j(\br)$ is the charge
density associated with the pair of zero modes $\chi_i$ and $\chi_j$. We observe
that at the neutrality point when $\tilde{K}_{ij}=0$  the low-energy
effective Hamiltonian (\ref{heff}) coincides with  the SYK
model. Eqs.\ (\ref{heffK}) and (\ref{heffJ}) allow us to calculate the
relevant two- and four-fermion coupling constants from the knowledge
of the Majorana wavefunctions in the non-interacting system. We shall
carry out this program in Section IV below for a specific physically
relevant model system. Here we finish by discussing some general
properties of Hamiltonian (\ref{heff}) that
follow from symmetry considerations.

The reality condition (\ref{real1}) for the Majorana wavefunction
implies the following spinor structure of $\Phi_j(\br)$ in the Nambu space  
\begin{equation}\label{real2}
\Phi_j=
\begin{pmatrix}
\eta_j \\
i\sigma^y\eta_j^*
\end{pmatrix},
\end{equation}
where $\eta_j(r)$ is a two-component complex spinor. We thus have
\begin{equation}\label{real3}
\rho_{ij}={i\over 2}(\eta_i^\dagger\eta_j-{\rm c.c.})=-{\rm
  Im}(\eta_i^\dagger\eta_j).
\end{equation}
 The charge density is thus purely real
and antisymmetric under $i\leftrightarrow j$. In the simplest case the
$\Sigma$-breaking part of the Fu-Kane Hamiltonian will simply be 
$\delta H_{\rm FK}(\br)=-\mu\tau^z$. In this situation Eq.\ (\ref{heffK})
implies that $\tilde{K}_{ij}=4\mu\int d^2r\rho_{ij}(\br)$. Thus $\tilde{K}_{ij}$ is
purely real and antisymmetric as required for ${\cal H}_{\rm eff}$ to
be hermitian.

Because of the anticommutation property (\ref{hsyk2}) of the Majorana operators it
is clear that only the fully antisymmetric part of $\tilde{J}_{ijkl}$
contributes to the Hamiltonian (\ref{heff}). As defined in
Eq.\ (\ref{heffJ}) $\tilde{J}_{ijkl}$ is already antisymmetric under
$i\leftrightarrow j$ and $k\leftrightarrow l$ due to the antisymmetry 
$\rho_{ij}=-\rho_{ji}$. With this in mind we can rewrite
Hamiltonian (\ref{heff}) in a more convenient form
\begin{equation}\label{heff2}
{\cal H}_{\rm eff}={i}\sum_{i<j} K_{ij}\chi_i\chi_j+\sum_{i<j<k<l} J_{ijkl}\chi_i\chi_j \chi_k\chi_l,
\end{equation}
with 
\begin{equation}\label{heff4}
K_{ij}={1\over 2}(\tilde{K}_{ij}-\tilde{K}_{ji}), \ \ \ 
{ J}_{ijkl}={1\over 3}(\tilde{J}_{ijkl}-\tilde{J}_{ikjl}+\tilde{J}_{lijk})
\end{equation}
now fully
antisymmetric. In the following we will be interested in situations
where coupling constants are random and will characterize the coupling
strengths by two parameters $K$ and $J$ defined by
\begin{equation}\label{heff3}
K^2=N \overline{K_{ij}^2},\ \ \ \ J^2={N^3\over 3!}\overline{J_{ijkl}^2},
\end{equation}
where the bar represents an ensemble average over randomness.

{
\subsection{Structure and statistics of the coupling constants $J_{ijkl}$}

In order to approximate the SYK Hamiltonian the coupling constants
$J_{ijkl}$ given in the previous subsection must behave as independent
random variables. To asess this condition we now discuss their
structure and statistics. We make two reasonable 
assumptions: (i) that the interaction potential in Eq.\ (\ref{heffJ})
is short ranged and well approximated by $V(\br)\simeq V_0\delta(\br)$,
and (ii) that there exists a lengthscale $\zeta$ beyond which Majorana
wavefunctions $\Phi_j(\br)$ can be treated as random independent
variables. 

We coarse-grain the
Majorana wavefunctions on the grid with with sites
$\br_n$  and spacing $\sim \zeta$.  This amounts to replacing
$\eta_j(\br)\to \bar\eta_j(\br_n)/\zeta$ and $\int d^2r
\to \zeta^2\sum_n$  in Eqs.\ (\ref{real2}) and (\ref{heffJ}). The discretized spinor wavefunctions then have the
following structure  on each site
\begin{equation}\label{eta1}
\bar\eta_j(\br_n)=
\begin{pmatrix}
\phi_j^1(n)+i\phi_j^2(n) \\
\phi_j^3(n)+i\phi_j^4(n) 
\end{pmatrix},
\end{equation}
where $\phi_j^\alpha(n)$ are real independent random variables with 
\begin{equation}\label{phi2}
\overline{\phi_i^\alpha(n)}=0, \ \ \ \overline{ \phi_i^\alpha(n)\phi_j^\beta(n)}={1\over 8M_s}\delta_{ij}\delta^{\alpha\beta}.
\end{equation}
Here $M_s=\pi R^2/\zeta^2$ is the total number of grid sites in the
hole and the second equality follows from the
normalization of $\Phi_j(\br)$.

Combining Eqs.\ (\ref{heffJ}), (\ref{real3}), (\ref{heff4}) and
(\ref{eta1}) it is possible to express the antisymmetrized coupling
constants as 
\begin{equation}\label{eta3}
J_{ijkl}=-{V_0\over \zeta^2}\sum_{n=1}^{M_s}\epsilon_{\alpha\beta\mu\nu} \phi_i^\alpha(n)\phi_j^\beta(n) \phi_k^\mu(n)\phi_l^\nu(n),
\end{equation}
}
{
where $\epsilon_{\alpha\beta\mu\nu}$ is the totally antisymmetric tensor and
summation over repeated indices is implied. For a general value of
$M_s$ the manybody Hamiltonian
defined by coupling constants Eq.\ (\ref{eta3}) represents a variant
of the original SYK model similar to models studied in Refs.\
\cite{Bi2017,Fu2016}. As such it might be amenable to the large-$N$
analysis using approaches described in those works. Here we focus on
the limit $M_s\gg N$ which attains when the hole radius $R$ is large
and the wavefunctions can be considered random on short scales $\zeta$. In
this limit  each $J_{ijkl}$ defined in Eq.\ (\ref{eta3}) is given by a
sum of a large number of random terms given by products of four random
amplitudes $\phi_j^\alpha(n)$. The central limit theorem then assures us that $J$'s
will be random variables with a distribution approaching the Gaussian
distribution irrespective of the detailed statistical properies of
$\phi_j^\alpha(n)$. It is furthermore easy to show that 
\begin{equation}\label{J2}
\overline{J_I}=0, \ \ \ \overline{ J_I J_J}=\left({3V_0\over
    8\zeta^2}\right)^2{1\over M^3_s}\delta_{IJ},
\end{equation}
where the uppercase label represents a group of four indices $I=\{ijkl\}$,
etc. The coupling constants given by Eq.\ (\ref{eta3}) are
asymptotically independent with the higher order correlators
vanishing as higher powers of $M_s$,
e.g.\ $\overline{J_{ijkl}J_{klmn}J_{mnij}}\sim M_s^{-5}$. 

The above analysis suggests that under reasonable assumptions coupling
constants defining the manybody Hamiltonian (\ref{heff2}) can be
considered independent random variables. When additionally $K_{ij}$
can be taken as negligible we expect the Hamiltonian to approximate
the SYK model.  Building on the experience gained from Refs.\
\cite{Bi2017,Fu2016} we furthermore expect our Hamiltonian to describe
an interesting non-Fermi liquid phase even away from the limit when
$J$'s are independent variables. For instance certain specific
correlations present in $J$'s are known to lead to a very interesting 
supersymmetric version of the SYK model \cite{Fu2016} and a whole
family of SYK-like models discussed in Ref.\ \cite{Bi2017} .

Recent work \cite{Alicea2017} performed a mathematical analysis of deviations in $J$'s from
ideal random independent variables in a model qualitatively similar to
ours. Here we adopt a different approach and proceed by evaluating the effect of such
deviations on the observable physical properties of the manybody
model defined by Hamiltonian (\ref{heff2}). We find that coupling constants that follow from the giant
vortex geometry indeed give rise to a phenomenology that is consistent with 
the SYK model. 
}

\section{The large-$N$ solution and the conformal limit}
When the number of Majorana fermions $N$ is large the SYK model
becomes analytically solvable in the low-energy limit. Specifically,
the Euclidean space time-ordered propagator defined as
\begin{equation}\label{prop1}
G(\tau)=\langle \cT_\tau\chi(\tau)\chi(0)\rangle
\end{equation}
can be expressed in the Matsubara frequency domain through the self energy
$\Sigma(\omega_n)$ as
\begin{equation}\label{prop2}
G(\omega_n)=[-i\omega_n-\Sigma(\omega_n)]^{-1}.
\end{equation}
Here $G(\omega_n)=\int_0^\beta d\tau e^{i\omega_n\tau}G(\tau)$ and
$\beta=1/k_BT$ is the inverse temperature. At non-zero temperatures
the propagator and the self energy are defined for discrete Matsubara
frequencies $\omega_n=\pi T(2n+1)$ with $n$ integer and taking
$k_B=1$ here and henceforth. 
Using the replica trick to average over disorder configurations,
or alternately summing the leading diagrams in the ${1\over N}$ expansion, 
one obtains (see for example Ref.\ \cite{Maldacena2016}) the following expression for the self energy appropriate
for  the Hamiltonian  (\ref{heff2})
\begin{equation}\label{prop3}
\Sigma(\tau)=K^2G(\tau)+J^2G^3(\tau).
\end{equation}

For arbitrary given parameters $K$, $J$ and $\beta$ the selfconsistent
Eqs. (\ref{prop2}) and (\ref{prop3}) can be solved by numerical
iteration. Analytical solutions are available in various limits and
will be reviewed below. In subsequent sections we shall compare these
with numerical results based on the model described above. 

\subsection{Free-fermion limit}
When $J=0$ the theory becomes noninteracting and an analytic solution to
Eqs. (\ref{prop2}) and (\ref{prop3}) can be given for all
temperatures. Specifically the self energy in Eq.\ (\ref{prop3}) can
be written in the
frequency domain as $\Sigma(\omega_n)=K^2 G(\omega_n)$ and substituted
into Eq.\ (\ref{prop2}). Solving for $G(\omega_n)$ then gives
\begin{equation}\label{prop4}
G_f(\omega_n)={2i\over \omega_n+\sgn(\omega_n)\sqrt{\omega_n^2+4K^2}}.
\end{equation}
This implies high-frequency limit $G_f(\omega_n)\simeq i/\omega_n$ and
low-frequency limit $G_f(\omega_n)\simeq i/\sgn(\omega_n)K$.

It is useful to extract the single-particle spectral function from
Eq.\ (\ref{prop4}) defined as $A(\omega)={1\over \pi}{\rm
  Im}G(\omega_n\to -i\omega +\delta)$,
by analytically continuing  from Matsubara to real frequencies to
obtain the retarded propagator.  We thus find
\begin{equation}\label{prop5}
A_f(\omega)={1\over \pi K}{\rm Re}\sqrt{1-\left({\omega\over 2K}\right)^2},
\end{equation}
the usual semicircle law. For this 0-dimensional system $A(\omega)$
coincides with the local density of states $D(\omega)$ averaged over all Majorana
sites which is experimentally measurable in a tunneling
experiment. Specifically, the tunneling conductance
$g(\omega)=(dI/dV)_{\omega=eV}$ is proportional to the local density
of states  $D(\omega)$. 
\begin{figure}[t]
\includegraphics[width = 7.5cm]{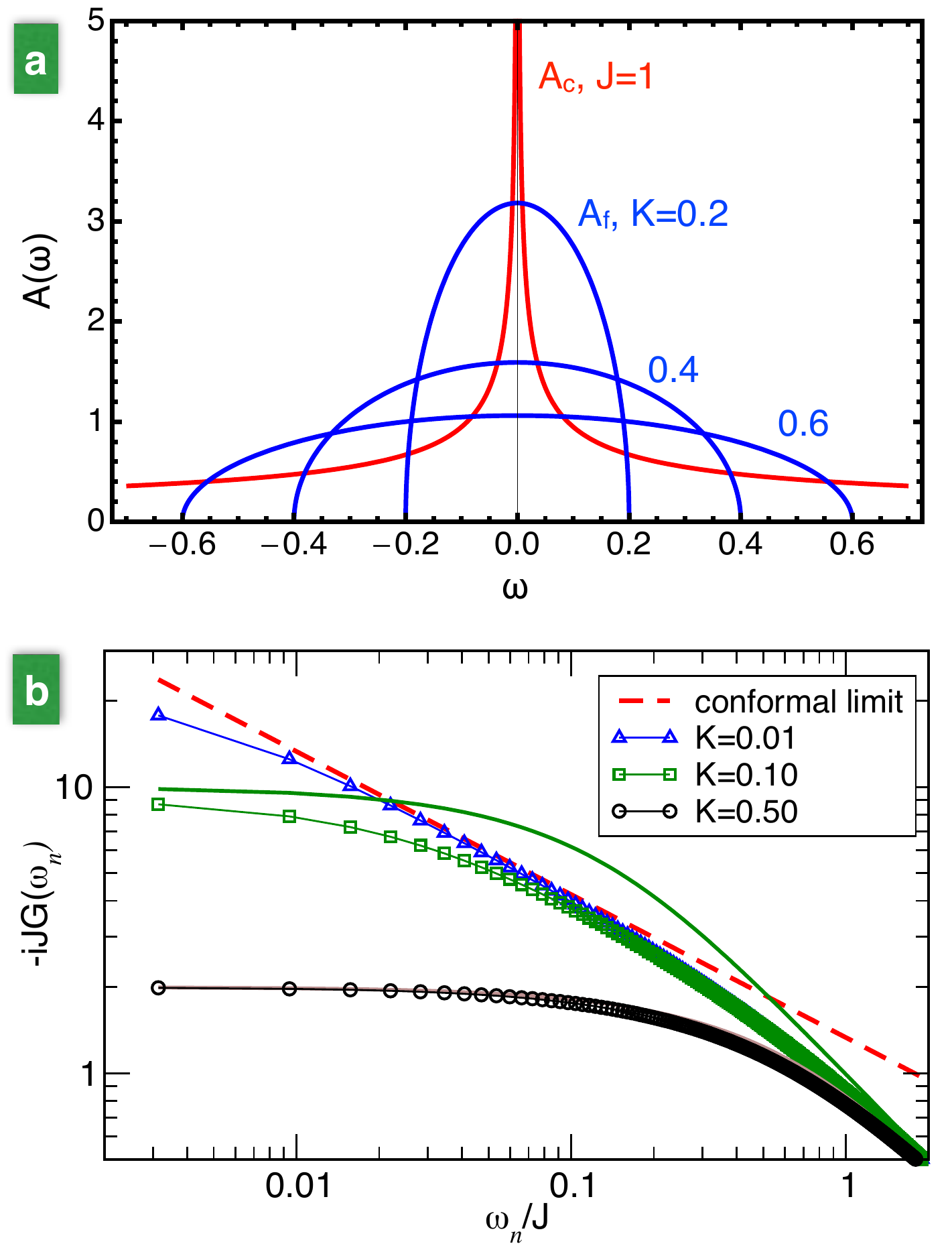}
\caption{a) Spectral functions, measurable in a tunneling experiment,
  in the conformal (strongly interacting) limit (red) and free-fermion
  limit (blue). b) Numerically evaluated large-$N$ Matsubara Green's
  functions for $J=1.0$, $T=0.001$ and different values of $K$. Red
  dashed line shows the conformal limit behavior Eq.\ (\ref{prop7})
  while the thick green and brown lines correspond to free-fermion
  result (\ref{prop4}) with $K=0.1$ and $0.5$, respectively.
}\label{fig2}
\end{figure}

\subsection{Conformal limit}
When $K=0$  and $T\ll J$ the system is strongly interacting but nevertheless
asymptotic solution of  Eqs. (\ref{prop2}) and (\ref{prop3}) can be
found by appealing to their approximate reparametrization invariance 
\cite{Kitaev2015,Maldacena2016}
that becomes exact in the low-frequency limit when one can neglect the
$-i\omega_n$ term in Eq.\  (\ref{prop2}). The conformal limit solution reads
\begin{equation}\label{prop6}
G_c(\omega_n)=i\pi^{1/4}{\sgn(\omega_n)\over\sqrt{J|\omega_n|}}
\end{equation}
and the corresponding spectral function 
\begin{equation}\label{prop7}
A_c(\omega)={1\over \sqrt{2}\pi^{3/4}}{1\over \sqrt{J|\omega|}}.
\end{equation}
These expressions are valid for $|\omega|\ll J$ and must cross over to
the $1/\omega$ behavior at large frequencies.  

It is important to note that the low-frequency behaviors of $A_f$ and
$A_c$ are quite different with the former saturating at $1/\pi K$
and the latter divergent. Thus it should be possible to distinguish the
free-fermion and the interaction dominated behaviors, illustrated in
Fig.\ \ref{fig2}a, by performing a tunneling experiment. We will discuss the measurement in more detail in Section \ref{sec:outlook}.

\subsection{Crossover region}
When both $K$ and $J$ are nonzero, as will be the case in a typical
experimental setup, analytical solutions are not available but one can
still understand the behavior of the system from approximate
analytical considerations and numerical solutions. Let us focus on the
$T=0$  limit and study the effect of $K$ and $J$ on the self energy
$\Sigma(\omega_n)$ in Eq.\ (\ref{prop3}). To this end it is useful to
consider the propagators $G_f$ and $G_c$ in the imaginary time
domain. For long times $\tau$ one obtains 
\begin{equation}\label{prop8}
G_f(\tau)={1\over \pi K}{\sgn(\tau)\over |\tau|}, \ \ \ 
G_c(\tau)={1\over \pi^{1/4} \sqrt{2J}}{\sgn(\tau)\over |\tau|^{1/2}}.
\end{equation}
Consider the $K=0$  limit and then slowly turn $K$ on. Initially,
$G_c(\tau)$ is a valid solution. However, for any non-zero $K$ it is
clear that the first term on the right-hand side of Eq.\ (\ref{prop3})
will dominate at sufficiently long times $\tau>\tau_*$. At such long
times one then expects a crossover to the  behavior resembling the
free-fermion propagator $G_f(\tau)$. The corresponding crossover time
$\tau_*$ can be estimated by equating the two terms on the right-hand
side  of Eq.\ (\ref{prop3}), $K^2G_f(\tau_*)=J^2G_c^3(\tau_*)$, which
gives
\begin{equation}\label{prop9}
\tau_*={\sqrt{\pi}\over 8}{J\over K^2},
\end{equation}
and the corresponding crossover frequency
\begin{equation}\label{prop10}
\omega_*={2\pi\over \tau_*}=16\sqrt{\pi}{K^2\over J}.
\end{equation}
We thus expect the spectral function to behave as indicated in Eq.\
(\ref{prop7}) for $\omega_*<\omega\ll J$ with the divergence at small
$\omega$ cut off below  $\omega_*$ and saturate to $\sim 1/\pi K$. 

To confirm the above behavior we have solved Eqs. (\ref{prop2}) and
(\ref{prop3}) numerically. We found it most convenient to work with
Matsubara Green's functions at very low but non-zero temperatures. To
this end we rewrite Eq.\  (\ref{prop3}) in Matsubara frequency domain where
the last term becomes a convolution and substitute the self energy
into Eq.\  (\ref{prop2}). We obtain a single equation
\begin{equation}\label{prop11}
G_n^{-1}=-i\omega_n-K^2G_n-J^2T^2\sum_{k,l}G_kG_lG_{n-k-l}
\end{equation}
for $G_n\equiv G(\omega_n)$ that must be solved
selfconsistently. Results obtained by iterating Eq.\ (\ref{prop11})
are displayed in Fig.\ \ref{fig2}b. For very small $K=0.01J$ we
observe that numerically calculated  $G(\omega_n)$ coincides with the
conformal limit for a range of frequencies consistent with our
discussion above. For $K=0.1J$ this range becomes smaller and
completely disappears for $K=0.5J$. 

We conclude that for any nonzero $K$ the
ultimate low-energy behavior is controlled by the free-fermion fixed
point, as expected on general grounds. Nevertheless, when $K$ is sufficiently small
in comparison to $J$, there can be a significant range of energies in
which the physics is dominated by the SYK fixed point. At low
temperatures the corresponding range of frequencies is given by 
\begin{equation}\label{prop12}
16\sqrt{\pi}{K^2\over J}<\omega\ll J.
\end{equation}
In this range we expect the spectral function to obey the conformal
scaling form given by Eq.\ (\ref{prop7}). Tunneling experiment in this
regime should therefore reveal the SYK behavior of the underlying
strongly interacting system.

\section{Numerical results: the underlying noninteracting system}

In this section we provide support for the ideas presented above by
performing extensive numerical simulation and modeling of the system
described in Sec.\ II.  We start by formulating a lattice model for
the surface of a TI in contact with a superconductor. We then find the
wavefunctions of the Majorana zero modes by numerically diagonalizing 
the corresponding Bogoliubov-de Gennes (BdG) Hamiltonian for the
geometry depicted in Fig.\ \ref{fig1} with $N$ flux quanta threading
the hole.  In the following Section, using Eqs. (\ref{heffK}) and (\ref{heffJ}), we
calculate the coupling constants $K_{ij}$ and $J_{ijkl}$, which we
then use to construct and diagonalize the many-body interacting
Hamiltonian  (\ref{heff2}) for $N$ up to 32. The resulting many-body
spectra and eigenvectors are used to calculate various physical
quantities (entropy, specific heat, two- and four-point propagators)
which are then compared to the results previously obtained for the SYK
model with random independent couplings.

\subsection{Lattice model for the TI surface}
A surface of a 3D TI is characterized by an odd number of massless
Dirac fermions protected by time reversal symmetry $\Theta$. The well known
Nielsen-Ninomyia theorem \cite{NIELSEN1981a,NIELSEN1981b} 
assures us that, as a matter of principle, it
is impossible to construct a purely 2D, $\Theta$-invariant lattice model with
an odd number of massless Dirac fermions. This fact causes a severe problem
for numerical approaches to 3D TIs because one is forced to perform an
expensive simulation of the 3D bulk to describe the anomalous 2D
surface. A workaround has been proposed \cite{Marchand2012} which
circumvents the Nielsen-Ninomyia theorem by simulating a pair of TI
surfaces with a total even number of Dirac fermions. This approach
enables efficient numerical simulations in a quasi-2D geometry while
fully respecting $\Theta$. 

\begin{figure}[t]
\includegraphics[width = 7.5cm]{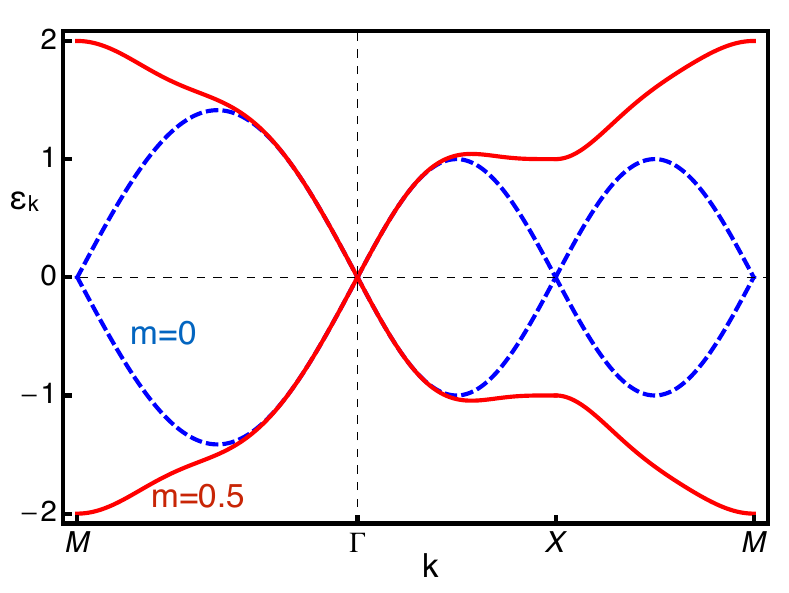}
\caption{Band structure (\ref{latt2}) of the lattice model Eq.\
  (\ref{latt1}) for $\lambda=1$ and $m=0$ (blue dashed) and $m=0.5$
  (red solid line). $X$ and $M$ denote the $(0,\pi)$ and $(\pi,\pi)$
  points of the Brillouin zone, respectively.
}\label{fig3}
\end{figure}
Here, because the physical time reversal symmetry is ultimately broken
by the presence of vortices and is therefore not instrumental, we opt for an even
simpler model which breaks $\Theta$ from the outset but nevertheless captures all
the essential physics of the TI/SC interface. We start from the
following momentum-space normal-state Hamiltonian defined on a simple 2D square
lattice   
\begin{equation}\label{latt1}
h_0(\bk)=\lambda(\sigma^x\sin{k_x}+\sigma^y\sin{k_y})+ \sigma^zM_\bk-\mu,
\end{equation}
with $M_\bk=m[(2-\cos{k_x}-\cos{k_y})-{1\over
  4}(2-\cos{2k_x}-\cos{2k_y})]$. Here ${\bm \sigma}$ are Pauli matrices in
spin space and $\lambda$, $m$ are model parameters. The term
proportional to $\lambda$ respects  $\Theta$ and gives 4 massless Dirac
fermions in accordance with the Nielsen-Ninomyia theorem. The $M_\bk$
term breaks $\Theta$ and has the effect of gapping out all the Dirac
fermions except the one located at $\Gamma=(0,0)$. The resulting
energy spectrum
\begin{equation}\label{latt2}
\varepsilon(\bk)=\pm\sqrt{\lambda^2(\sin^2{k_x}+\sin^2{k_y})+ M^2_\bk}-\mu,
\end{equation}
is depicted in Fig.\ \ref{fig3}. In the vicinity of the $\Gamma$ point
we observe a linerly dispersing spectrum characteristic of  a TI surface state. It is  to be noted that for small
$|\bk|$ we have $M_\bk\simeq {1\over 8}mk^4$ so the amount of
$\Theta$-breaking can be considered small in the physically important
part of the momentum space near the $\Gamma$ point.

Proximity induced superconducting order is implemented by constructing
the BdG Hamiltonian, 
\begin{equation}\label{latt3}
H_{\rm BdG}(\bk) =
\begin{pmatrix}
h_0(\bk) & \Delta\\
\Delta^* & -\sigma^yh_0^*(-\bk)\sigma^y
\end{pmatrix}.
\end{equation}
Writing $H_{\rm BdG}$ in terms of $\sigma$ and $\tau$ matrices it can
be easily checked that it respects the particle-hole symmetry $\Xi$
defined in Sec.\ II.A.  The $\mu$ and $M_\bk$ terms both break the
fictitious time reversal $\Sigma$ that protects the Majorana zero
modes in our setup. As before $\mu$ must be tuned to zero to achieve
protection. 
{ On the other hand it is crucial to remember that
$M_\bk$ has been introduced only to circumvent the  Nielsen-Ninomyia
theorem and allow us to efficiently simulate a single two-dimensional 
Dirac fermion on the lattice.  Breaking of $\Sigma$ by $M_\bk$ is therefore not a
concern in the experimental setup: in a real TI tuned to the neutrality
point $\Sigma$ is unbroken. Expanding $H_{\rm BdG}(\bk)$ in the
vicinity of $\Gamma$ to leading order in small $\bk$ we recover
the Fu-Kane Hamiltonian $H_{\rm FK}$ defined in Eq.\ (\ref{sym1}). We
thus conclude that at low energies our lattice model indeed describes the
TI/SC interface and should exhibit the desired phenomenology,
including Majorana zero modes bound to vortices. We shall see that
this is indeed the case.  The
only repercussion that follows from the weakly broken $\Sigma$
(present in the  higher order
terms in the above expansion) is 
a very small splitting of the zero mode energies that has no
significant effect on our results.}

\subsection{Solution in the giant vortex geometry}

\begin{figure*}[t]
\includegraphics[width = 14.5cm]{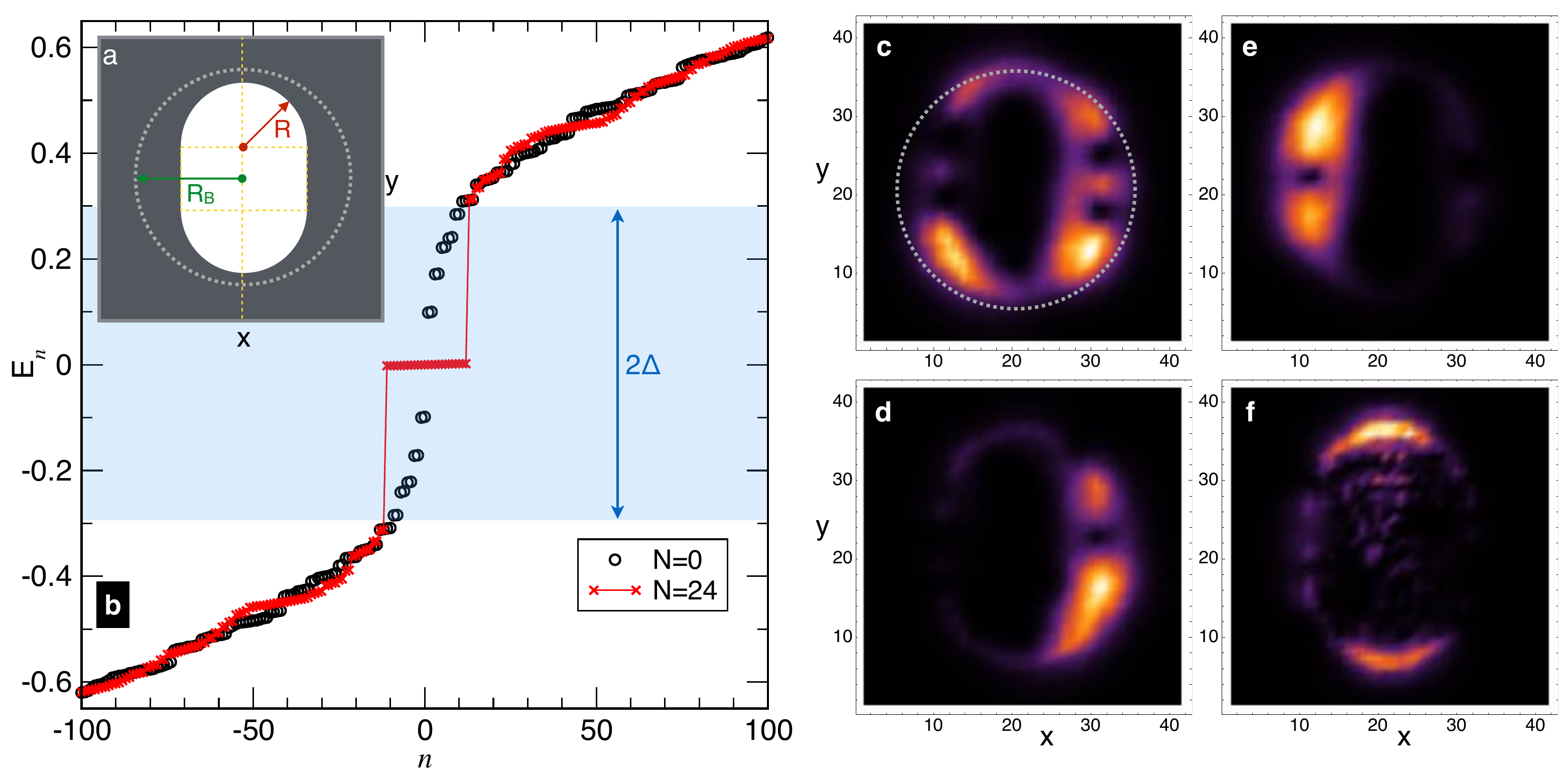}
\caption{Numerical simulations of the BdG Hamiltonian Eq.\
  (\ref{latt5}). a) Stadium-shaped hole geometry employed in the
  simulations. $R$ parametrizes the hole size whereas $R_B$ denotes
  the radius inside which the magnetic field is nonzero. 
b) Energy levels $E_n$ of the BdG Hamiltonian 
  (\ref{latt5}) calculated for $N=0$ and $N=24$. Energies have been
  sorted in ascending order and plotted as a function
  of their integer index $n$. The shaded band represents the SC gap region. c-f) Density plots of
  the typical zero mode wavefunction amplitudes for $N=24$. The dashed circle in panel (c)
  has radius $R_B$. The following
  parameters have been used to obtain these results: $\lambda=1$,
  $m=0.5$, $\Delta_0=0.3$, $\mu=w_\mu=0$, $w_\lambda=w_\Delta=0.1$,
  $L=42$, $R=10$ and $R_B=15$. 
}\label{fig4}
\end{figure*}
To study the non-uniform system with magnetic field and vortices we
must write the Hamiltonian in the position space.  The normal-state
piece Eq.\ (\ref{latt1}) is most conveniently written in
second-quantized form as
\begin{eqnarray}
\cH_0&=&
     i\lambda\sum_{\br,\alpha}\left(\psi_\br^\dagger\sigma^\alpha\psi_{\br+\alpha}-{\rm
     h.c.}\right) +\sum_\br \psi_\br^\dagger\left({3\over 2}m\sigma^z-\mu\right)\psi_\br
     \nonumber \\
&-&{m\over
    8}\sum_{\br,\alpha}\left(4\psi_\br^\dagger\sigma^z\psi_{\br+\alpha}-
\psi_\br^\dagger\sigma^z\psi_{\br+2\alpha}+{\rm h.c.}\right) \label{latt4},
\end{eqnarray}
where we defined on each lattice site $\br$ a two-component spinor
$\psi_\br=(c_{\br\uparrow},c_{\br\downarrow})^T$ and $\alpha=x,y$. The
magnetic field is included through the standard Peierls substitution
which replaces  tunneling amplitudes on all bonds according to
$\psi_\br^\dagger\psi_{\br+\alpha}\to\psi_\br^\dagger\psi_{\br+\alpha}\exp(-i{e\over \hbar c}\int_\br^{\br+\alpha}d{\bm l}\cdot\bA)$.
The full second quantized BdG Hamiltonian then reads
\begin{equation}\label{latt5}
\cH_{\rm BdG}=\cH_0+\sum_\br(\Delta_\br c^\dagger_{\br\uparrow}c^\dagger_{\br\downarrow}+{\rm h.c.}),
\end{equation}
where $\Delta_\br$ is the pair potential on site $\br$ which takes
the form indicated in Eq.\ (\ref{gap1}).
{ In accord with our discussion in the previous subsection
 $\cH_{\rm BdG}$ given in Eq.\ (\ref{latt5}) represents a version of
 the Fu-Kane Hamiltonian (\ref{abel1}) regularized on a square
 lattice. This lattice model is suitable for numerical calculations
 and we expect it to reproduce all the low-energy features of the
 Fu-Kane Hamiltonian. In particular we will see shortly that it yields
 $N$ Majorana zero modes mandated by the Jackiw-Rossi-Weinberg
 index theorem that are of central importance for the SYK model.}

It is most convenient to solve the problem defined by Hamiltonian
(\ref{latt5}) on a lattice with $L\times L$ sites and periodic
boundary conditions which ensure that no spurious edge states exist at
low energies in addition to the expected $N$ Majorana zero modes bound
to the hole. To implement periodic boundary conditions it is useful to perform a singular gauge
transformation 
\begin{equation}\label{latt6}
\psi_\br\to e^{iN\varphi/2}\psi_\br ,
\end{equation}
which has the effect of removing the phase winding from $\Delta_\br$
and changing the Peierls phase factors to
$\exp(i\int_\br^{\br+\alpha}d{\bm l}\cdot{\bm \Omega})$ with 
\begin{equation}\label{latt7}
{\bm \Omega}={1\over
  2}\left(N\nabla\varphi-{2e\over \hbar c}\bA\right).
\end{equation}
 We note that $N$ must be
even because only for integer number of fundamental flux quanta
$hc/e=2\Phi_0$ in the system  one can impose periodic boundary conditions.
For $N$ even the transformation (\ref{latt6}) is single valued and the issue of
branch cuts that renders the analogous problem with singly-quantized vortices
\cite{FT2000,FT2001} more complicated does not arise here.
After the transformation the total effective flux seen by the
electrons $\int dS(\nabla\times {\bm \Omega})_z$ vanishes and
numerical diagonalization of the transformed Hamiltonian
(\ref{latt5}) with periodic boundary conditions becomes
straightforward. 

As a practical matter it is easiest to define a
regular shaped hole and introduce disorder through a replacement 
\begin{equation}
(\mu,\lambda,\Delta_\br)\to (\mu,\lambda,\Delta_\br)+
(\delta\mu_\br,\delta\lambda_\br,\delta\Delta_\br) \label{latt8}
\end{equation}
Here $(\delta\mu_\br,\delta\Delta_\br,\delta\lambda_\br)$ are independent
random variables uniformly distributed in the interval $(-w_\mu/2,w_\mu/2)$ for  $\delta\mu_\br$
and similarly for $\delta\Delta_\br$ and $\delta\lambda_\br$. We chose a stadium-shaped hole
sketched in Fig.\ \ref{fig4}a  which is known to support classically
chaotic trajectories \cite{gutzwiller1990chaos, haake2013quantum}. In our
quantum simulation we find that much smaller disorder strength is
required to achieve sufficiently random Majorana wavefunctions for
stadium-shaped hole than e.g.\ a with circular hole. We furthermore chose
magnetic field $B$ to be uniform inside the radius $R_B$ that contains
the hole and zero otherwise. We find that our results are insensitive
to the detailed distribution of $B$ as long as the total flux remains
$N\Phi_0$ and is centered around the hole  (we tested various radii
$R_B$ as well as a Gaussian profile).

Typical results of the numerical simulations described above are
displayed in Fig.\ \ref{fig4}.  In panel (b) we observe the behavior
of the energy eigenvalues $E_n$ of $\cH_{\rm BdG}$. For zero magnetic
flux  there are several states inside the SC gap (Andreev states bound
to the hole) but no zero modes. For $N=24$ these are converted into 24
zero modes required by the Jackiw-Rossi-Weinberg index theorem. For
$\mu=w_\mu=0$ used in the simulation their energies are very close to
zero ($\sim 10^{-4}\lambda$), where the small residual splitting is
attributable to the fact that $\Sigma$ symmetry is weakly
broken in our lattice simulation by the $M_\bk$ term. For non-zero
$\mu$ or $w_\mu$ the energy splitting increases in proportion to these
$\Sigma$-breaking perturbations. In the following we include
these terms in $\delta\cH_{\rm FK}$ and incorporate them in our
many-body calculation via $K_{ij}$ terms given by Eq.\ (\ref{heffK}).

 Panels Fig.\ \ref{fig4} c-f
show examples of zero mode wavefunction amplitudes $|\Phi_j(\br)|^2$. The wavefunctions
are seen to exhibit random spatial structure which depends sensitively
on the specific disorder potential realization. Importantly all the
zero mode wavefunctions are localized in the same region of space
defined by the hole and its immediate vicinity. One therefore expects
Eq.\ (\ref{heffJ}) to produce strong random couplings $J_{ijkl}$
connecting all zero modes $\chi_j$ once the interactions are included.

\section{Numerical results: the many-body SYK problem}

Having obtained the zero mode wavefunctions it is straightforward to
calculate couplings $K_{ij}$ and $J_{ijkl}$ from Eqs.\ (\ref{heffK})
and (\ref{heffJ}) and construct the many-body SYK Hamiltonian 
(\ref{heff2}). In the following we shall assume that the system has
been tuned to its global neutrality point $\mu=0$ and include in
$\delta\cH_{\rm FK}$ only the random component of the on-site
potential $\delta\mu_\br$. For the interaction term we consider the screened
Coulomb potential defined as 
\begin{equation}\label{syk1}
V(r)={2\pi e^2\over \epsilon}{e^{-r/\lambda_{\rm TF}}\over r},
\end{equation}
where $\epsilon$ is the dielectric constant and $\lambda_{\rm TF}$
denotes the Thomas-Fermi screening length. We furthermore assume that
$\lambda_{\rm TF}$ is short so that in the lattice model the
interaction is essentially on-site. The expression for
$\tilde{J}_{ijkl}$ then simplifies to 
\begin{equation}\label{syk2}
\tilde{J}_{ijkl}\simeq 12V_0\int d^2r \rho_{ij}(\br)\rho_{lk}(\br),  
\end{equation}
with $V_0=\int d^2r V(r)= 2\pi e^2 \lambda_{\rm
  TF}/\epsilon$. Coupling constants $K_{ij}$ and $J_{ijkl}$ are
easy to evaluate using Eq.\ (\ref{heff4}) and the Majorana wavefunctions $\Phi_j(\br)$
obtained in the previous Section.  To facilitate comparisons with the
existing literature we shall quantify the average strength of these terms
using parameters $K$ and $J$ defined in Eq.\
(\ref{heff3}). Specifically  we will adjust $w_\mu$ and $V_0$ to obtain the desired values
of $K$ and $J$. In the following Section we will connect these values
to the parameters expected in realistic physical systems.

\subsection{Thermodynamic properties and many-body level statistics}

\begin{figure}[t]
\includegraphics[width = 7.5cm]{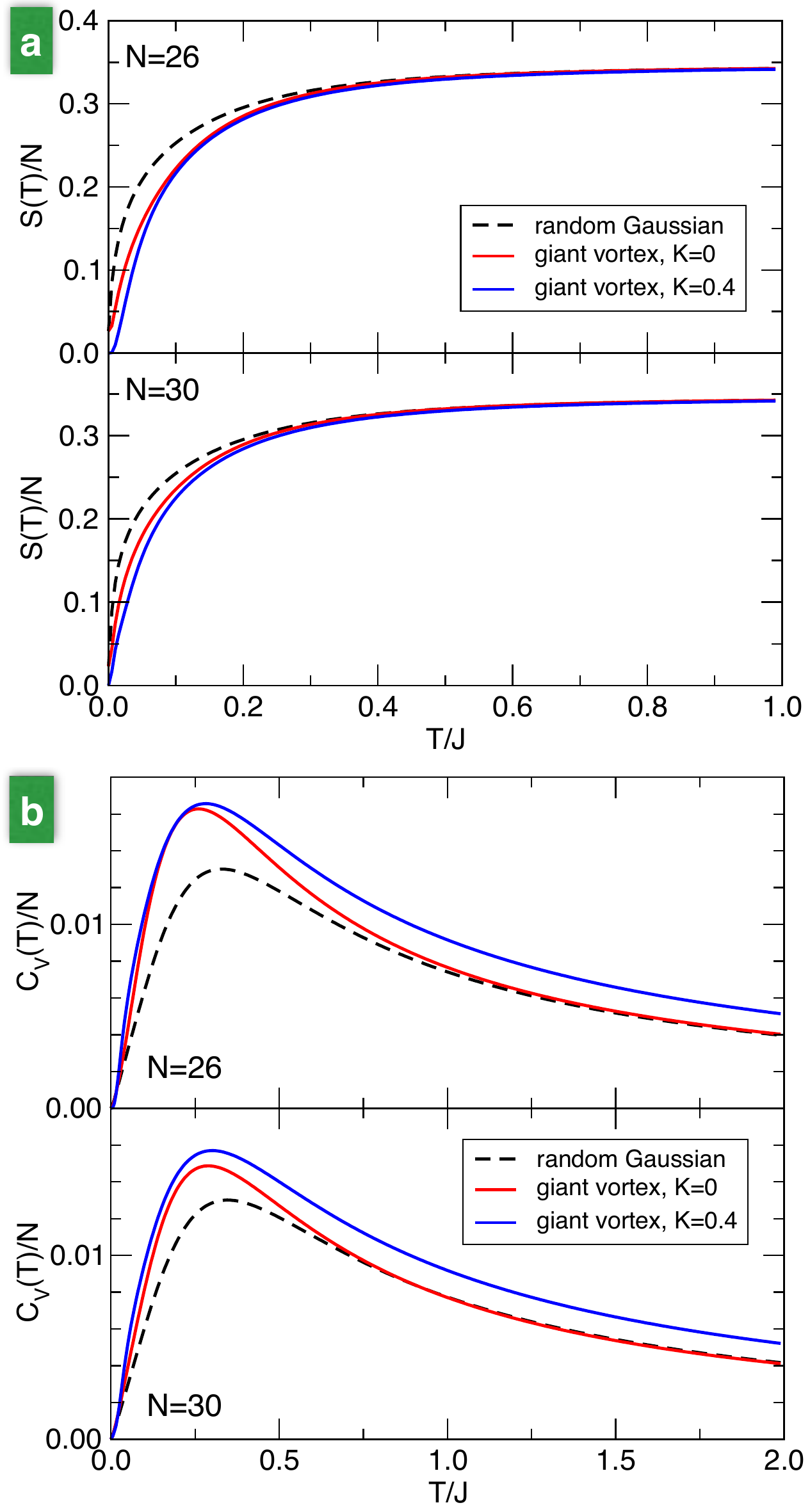}
\caption{Thermodynamic properties of the many-body Hamiltonian
  (\ref{heff2}), a) thermal entropy per particle and b) heat capacity
  per particle. Dashed lines show the expected behavior for the SYK
  model with random independent couplings, solid lines show results
  for the couplings obtained from the giant vortex system. In all
  panels the same parameters  have been used as in Fig.\ \ref{fig4}
  with $V_0$ adjusted so that $J=1$.   
}\label{fig5}
\end{figure}
Once the coupling constants $K_{ij}$ and $J_{ijkl}$ are determined as
described above one can construct a matrix representation of the many-body
Majorana Hamiltonian (\ref{heff2}) and find its energy eigenvalues
$E_n$ by exact numerical diagonalization. From the knowledge of the
energy levels it is straightforward to calculate any thermodynamic property. In
Fig.\ \ref{fig5} we display the thermal entropy $S(T)$ and
the heat capacity  $C_V(T)$.  These are calculated from 
\begin{equation}\label{syk3}
S={\langle E\rangle -F\over T}, \ \ \ \ C_V={\langle E^2\rangle -\langle E\rangle^2\over T},
\end{equation}
where $\langle E^\alpha\rangle={1\over Z}\sum_n E_n^\alpha e^{-E_n/T}$,
$F=-T\ln{Z}$ is the free energy and $Z=\sum_n e^{-E_n/T}$ the
partition function. 

The entropy per particle is seen to saturate at
high temperature to $S_\infty/N={1\over 2}\ln{2}\simeq 0.3465$ as expected for a
system of $N$ Majorana fermions. The behavior of $S(T)$
calculated for the giant vortex system is qualitatively similar to
that obtained from the
SYK model with random independent couplings. The small deviations
that exist are clearly becoming smaller as $N$ grows, suggesting that
they vanish in the thermodynamic limit. Nonzero two-body coupling $K$
is seen to modify the entropy slightly at low temperature. For large
$N$ and $K=0$ the entropy per particle is expected to attain a nonzero
value $\sim 0.24$ as $T\to 0$ due to the extensive ground-state degeneracy of the SYK
model. Our largest system is not large enough to show this behavior
(in agreement with previous numerical results) 
although Fig.\ \ref{fig5}a correctly captures the expected suppression
of the low-$T$ entropy in the presence of  two-body couplings which
tend to remove the  extensive ground-state degeneracy.

The heat capacity $C_V(T)$, displayed in Fig.\ \ref{fig5}b, likewise
behaves as expected for the SYK model with random independent
couplings with small deviations becoming negligible in the large-$N$
limit. $C_V(T)$ is in principle measurable and we can see from Fig.\
\ref{fig5}b that its high-temperature behavior could be used gauge the
relative strength of two- and four-fermion terms in the system.

\begin{figure*}[t]
\includegraphics[width = 17.5cm]{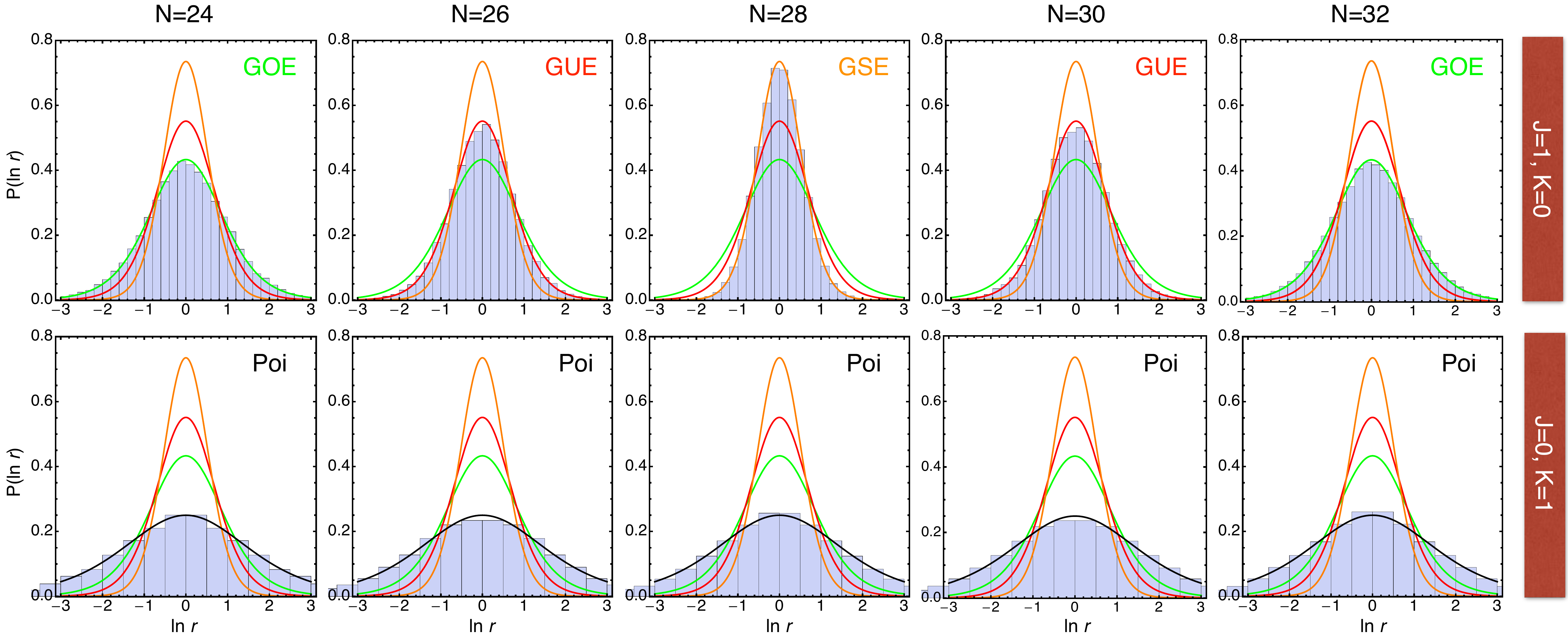}
\caption{Level statistics analysis.  Top row shows histograms of
  $\ln{r_n}$ obtained from the energy levels of the SYK Hamiltonian
  (\ref{heff2}) with coupling constants taken from the giant vortex
  model with $V_0$ and $w_\mu$ adjusted so that $J=1$ and $K=0$. Solid
  lines indicate the expected distributions GOE (green), GUE (red) and
  GSE (orange) specified in Eq.\ (\ref{syk4}).  Bottom row shows
  results for the noninteracting case $J=0$, $K=1$. Black solid line
  represents the Poisson distribution Eq.\ (\ref{syk5}). Histograms in all
  panels have been averaged over 8 independent realizations of
  disorder except for $N=24$ where 16 realizations have been employed
  to obtain satisfactory statistics and $N=32$ for which a single
  realization was used.
}\label{fig6}
\end{figure*}
As discussed in Refs.\ \cite{Verbaar2016,Xu2016} many-body level statistics provides a sensitive
diagnostic for the SYK physics encoded in the Hamiltonian
(\ref{heff2}). To apply this analysis to our results we arrange the
many-body energy levels in ascending  order $E_1<E_2< \dots$ and form
 ratios between the successive energy spacings
\begin{equation}\label{sykr}
 r_n={E_{n+1}-E_n\over E_{n+2}-E_{n+1}}. 
\end{equation}
 According to Refs.\
 \cite{Verbaar2016,Xu2016} the SYK Hamiltonian can be constructed as a
 symmetric matrix in the Clifford algebra ${\cal C}\ell_{0,N-1}$ whose
 Bott periodicity gives rise to a Z$_8$ classification with
 topological index $\nu=N\!\mod{8}$. As a result 
 statistical distributions of the ratios $P(r)$ cycle through
 Wigner-Dyson random matrix ensembles with Z$_8$ periodicity as a
 function of $N$. Specifically, 
 Gaussian orthogonal (GOE), unitary (GUE) and symplectic (GSE)
 ensembles occur with distributions given by the ``Wigner surmise''
\begin{equation}\label{syk4}
P(r)={1\over Z}{(r+r^2)^\beta\over (1+r+r^2)^{1+3\beta/2}}
\end{equation}
and parameters $Z$ and $\beta$ summarized in Table I for even $N$ relevant to our system.
\begin{table}\label{ttable1}
\begin{tabular*}{0.48\textwidth}{@{\extracolsep{\fill}}l | c c c c }
\hline \hline 
 $N(\!\!\mod{8})$  & 0  & 2 & 4 & 6  \\
\hline
level stat. & GOE & GUE & GSE & GUE \\
$\beta$ & 1 & 2 & 4 & 2 \\
$Z$ & ${8\over 27}$ & ${4\pi\over 81 \sqrt{3}}$ & ${4\pi\over
                                                  729\sqrt{3}}$  &
                                                                   ${4\pi\over 81 \sqrt{3}}$ \\
\vspace{-4pt}\\
\hline \hline
\end{tabular*}
\caption{Gaussian ensembles for even $N$.}
\end{table}
As emphasized in Ref.\ \cite{Xu2016} the level spacing analysis must
be performed separately in the two fermion parity sectors of the
Hamiltonian  (\ref{heff2}).

Fig.\ \ref{fig6} shows statistical distributions of the ratios $r_n$
computed for $N=24,26,28,30$ and 32 in our system. For the sake of clarity
$P(\ln{r})$ is plotted along with the anticipated distributions for
GOE, GUE and GSE given in Eq.\ (\ref{syk4}). Unambiguous agreement
with the pattern indicated in Table I is observed, lending
further support to the notion that our proposed system realizes the
SYK model. We checked that the Z$_8$ periodic pattern persists for all
$N$ down to 16. Additionally, the above results should be contrasted with
the level statistics in the noninteracting case $J=0$, $K=1$ displayed
in the bottom row of Fig.\ \ref{fig6}. In the absence of interactions
Z$_8$ periodicity is absent and the distribution of the ratio $r_n$
follows Poisson level statistics
\begin{equation}\label{syk5}
P(r)={1\over (1+r)^2}
\end{equation}
for all $N$. It is to be noted that no adjustable parameters are
employed in the level-statistics analysis presented above.

\subsection{Green's function}
\begin{figure}[t]
\includegraphics[width = 8.5cm]{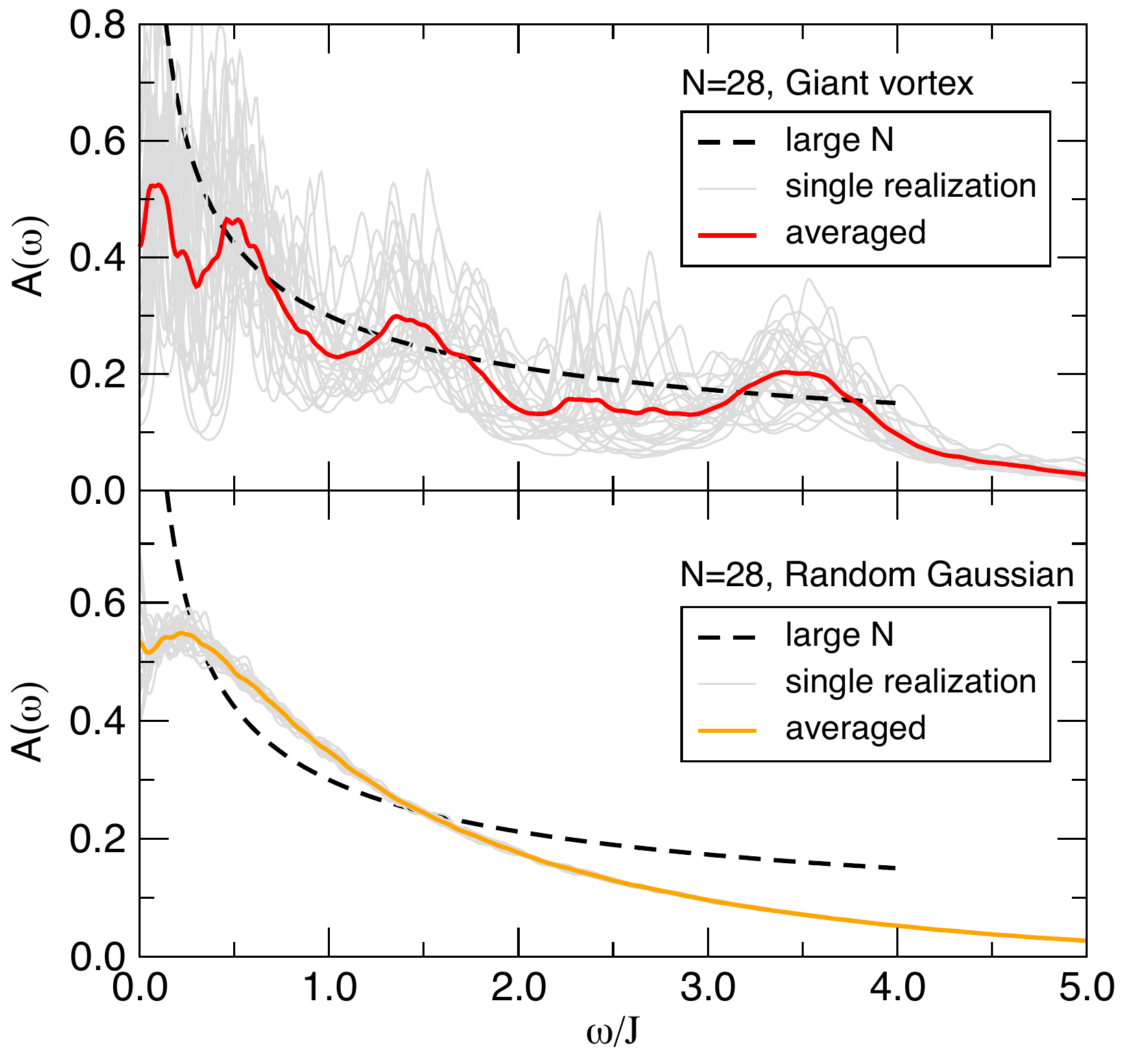}
\caption{Spectral function $A(\omega)$ computed at zero temperature
for coupling constants $J_{ijkl}$  obtained from the giant vortex
calculation (top) and taken from the Gaussian distribution (bottom). 
Thin grey lines represent  individual disorder realizations
corresponding to a physical measurement in a system with
quenched disorder. Thick lines reflect the average over 25 independent disorder
realizations. Dashed lines represent the expected low-frequency
behavior in the large-$N$ conformal limit Eq.\ (\ref{prop7}). All
parameters are as in Fig.\ \ref{fig4} with $J=1$,  $K=0$ and
broadening $\delta=0.04$ in Eq.\ (\ref{syk7}). 
}\label{fig7}
\end{figure}
Computing the Green's function of the model is perhaps the most
straightforward way of comparing the behavior of the system at finite
$N$ to the large-$N$ limit solutions discussed in Sec.\ III. At the
same time computation of propagators is numerically more costly
because in addition to many-body energy levels one requires the
corresponding eigenstates. We computed the on-site retarded Green's
function defined as
\begin{equation}\label{syk6}
G_i^R(t-t')=-i\theta(t-t')\langle\{\chi_i(t),\chi_i(t')\}\rangle.
\end{equation}
Fourier transforming and using Lehmann representation in terms of the eigenstates
$|n\rangle$ of the many-body Hamiltonian (\ref{heff2}) one obtains, at
$T=0$,
\begin{equation}\label{syk7}
G_i^R(\omega)=\sum_n\left[{|\langle n|\chi_i|0\rangle|^2\over
    \omega+E_0-E_n+i\delta}
+(E_0\leftrightarrow E_n)\right],
\end{equation}
where $\delta$ is a positive infinitesimal. From Eq.\ (\ref{syk7}) the spectral
function $A_i(\omega)={1\over \pi}{\rm Im}G_i^R(\omega)$ is readily
extracted.

In Fig.\ \ref{fig7} we display spectral function $A(\omega)={1\over
N}\sum_iA_i(\omega)$ averaged over all Majorana zero modes. Physically
this corresponds to a tunneling experiment with a large probe that
allows for tunneling into all sites inside the hole. In agreement with
the existing numerical results on the complex fermion version of the
SYK model \cite{Wengbo2016} we find that for system sizes we can
numerically access (up to $N=30$) the conformal limit is approached
only in a narrow interval of frequencies. In the low-frequency limit
numerical results approach a constant value instead of the the 
$\sim 1/\sqrt{\omega}$ divergence expected in the large-$N$ limit. 
The dependence on $N$ is very weak,
with the larger values showing reduced statistical fluctuations but otherwise
qualitatively similar behavior. To convincingly demonstrate the conformal scaling of the
Green's function at the  lowest frequencies numerical calculations using larger
values of $N$ would be necessary. Unfortunately these are currently
out of reach for the exact diagonalization method due to the
exponential growth of the Hamiltonian matrix size with $N$. More
sophisticated numerical techniques, such as the quantum Monte Carlo,
could possibly reach larger system sizes.
 
Spectral functions calculated for the giant
vortex setup exhibit larger statistical fluctuations compared to those
computed  with random Gaussian coupling constants
$J_{ijkl}$ but are qualitatively similar when averaged over
independent disorder realizations.
 Therefore, we conclude that the Green's function behavior at
finite $N$ supports the notion that our proposed system realizes the
SYK model.

\subsection{Out-of-time-order correlators and scrambling}
Scrambling of quantum information  -- a process in which quantum
information deposited into the system locally gets distributed among
all its degrees of freedom -- is central to the conjectured duality
between the SYK model and AdS$_2$ Einstein gravity. Black holes are
thought to scramble with the maximum possible efficiency: they exhibit
quantum chaos. For a quantum theory to be the holographic dual of a
black hole its dynamics must exhibit similar fast scrambling
behavior. 
\begin{figure}[t]
\includegraphics[width = 8.5cm]{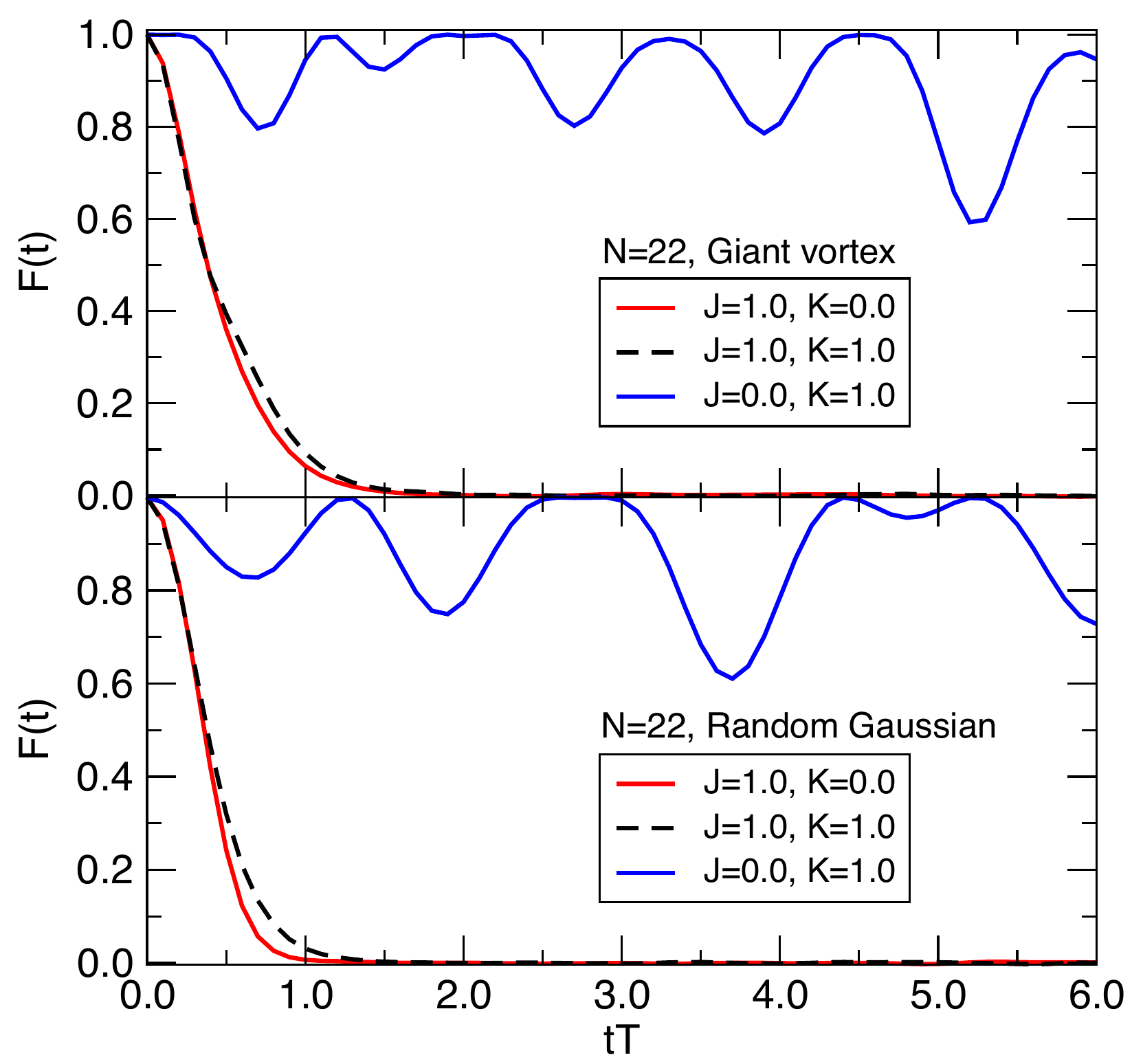}
\caption{Out-of-time-order correlators for our giant vortex system
  (top) and random Gaussian couplings (bottom). Non-zero temperature
  $T=1$ has been taken and other parameters as in Fig.\
  \ref{fig4}. Correlators are displayed for a single disorder
  realization (unaveraged) but different realizations give very
  similar results for all interacting cases. Detailed shape of the oscillations apparent in
  the $J=0$ curves depends sensitively on the specific disorder realization
  but all realizations show qualitatively similar behavior.
}\label{fig8}
\end{figure}

Out-of-time-order correlator (OTOC), defined in our system as 
\begin{equation}\label{syk10}
F_{ij}(t)=\langle\chi_j(t)\chi_i(0)\chi_j(t)\chi_i(0)\rangle,
\end{equation}
allows to quantify the quantum chaotic behavior. For black holes in
Einstein gravity scrambling occurs exponentially fast with $1-F(t)\sim
e^{\lambda_L t}/N$ where the decay rate is given by the Lyapunov
exponent $\lambda_L=2\pi T$ \cite{Maldacena2016b}. Similarly, for the SYK model in the
large-$N$ limit one expects \cite{Kitaev2015,Maldacena2016}
\begin{equation}\label{syk11}
1-F(t)\sim {J\over NT}e^{\lambda_Lt}.
\end{equation}
Previous works \cite{Hosur2016,Wengbo2016} gave numerical evaluations
of $F(t)$ in the SYK model for $N$ up to 14 but found these system sizes
to be too small to clearly show the expected $J$-independent Lyapunov
exponent. Here we numerically evaluate OTOC for $N$ up to 22 and show
that coupling constants obtained from the giant vortex geometry give
qualitatively the same behavior as those for random independent
coupling constants. Our results are summarized in Fig.\ \ref{fig8}
where we compute the on-site OTOC $F_{ii}(t)$ averaged over all sites.

For $J=1$ the OTOC is seen to rapidly decay to zero consistent with previous
works on the SYK model \cite{Hosur2016,Wengbo2016}. The rate of decay is controlled by
$J$: as in Refs.\ \cite{Hosur2016,Wengbo2016} we find that $N=22$ is not
large enough to observe  the theoretically predicted $J$-independent Lyapunov
exponent controlled by temperature, even when $J\gg T$. In addition we
observe that adding sizable two-body tunneling term $K$ has only very
modest effect on the behavior of $F(t)$ when the interaction strength
is maintained. However, in the non-interacting case  $(J=0,\  K=1)$ OTOC
behavior changes qualitatively with the fast decay replaced by
oscillations whose amplitude slowly increases.

\section{Outlook: Towards the experimental realization and detection of the SYK model \label{sec:outlook}}

Our theoretical results presented above indicate that low-energy
fermionic degrees of freedom in a device with
geometry depicted in Fig.\ \ref{fig1} provide a physical realization
of the SYK model. Additionally, all the ingredients are currently in place to
begin experimental explorations of the proposed
system. Superconducting order has been induced and observed at the
surface of several TI compounds by multiple groups
\cite{Koren2011,Sacepe2011,Qu2012,Williams2012,Cho2013,Xu2014,Zhao2015}. 
{ 
Importantly, Ref.\  \cite{Cho2013} already demonstrated
the ability to tune the chemical potential in
(Bi$_x$Sb$_{2-x}$)Se$_3$ thin flakes through the neutrality point {\em in
the presence of superconductivity} induced by Ti/Al contacts  by a
combination of chemical doping (tuning $x$) and back gate
voltage. This is almost exactly what we require for the implementation
of the SYK model.}
Well
developed techniques (such as focused ion milling)  exist to fabricate patterns, such as a
hole with an irregular shape, in a SC  film deposited on the TI
surface. In the remainder of this Section we discuss in more detail
the experimentally relevant constraints on the proposed device as well
as possible ways to detect manifestations of the SYK physics in a
realistic setting.

\subsection{Device geometry, length and energy scales}    
The key controllable design feature is the size of the hole,
parametrized by its radius $R$. For simplicity in the estimates below we shall assume a
circular hole but it should be understood that in a real experiment
irregular shape is required to promote randomness of the zero mode
wavefunctions. For the desired number $N$ of Majorana zero modes the
hole must be large enough to pin $N$ vortices. Vortex pinning occurs
because the SC order parameter $\Delta$ is suppressed to zero in the
vortex core which costs condensation energy. Vortices therefore prefer
to occupy regions where $\Delta$ has been locally suppressed by
defects or in
our case by an artificially fabricated hole. The optimal hole size $R_N$  for
$N$ vortices in our setup can be thus estimated from the requirement
that all the electronic states inside the hole that reside within the
SC gap are transformed into zero modes,
\begin{equation}\label{exp1}
\pi R_N^2\int_{-\Delta}^\Delta d\varepsilon D(\varepsilon) = N,
\end{equation}
where $D(\varepsilon)=|\varepsilon|/2\pi v_F^2\hbar^2$ is the density of
states of the TI surface. This gives
\begin{equation}\label{exp2}
 R_N=\pi\xi\sqrt{2N},
\end{equation}
with $\xi=\hbar v_F/\pi\Delta$ the BCS coherence length. In the
absence of interactions a hole of this size will produce an energy
spectrum similar to that depicted in  Fig. \ref{fig4}b, with $N$ zero
modes maximally separated from the rest of the spectrum. 

{
In reality, if the SC film is in the type-II regime,
a somewhat larger hole might be required to reliably pin $N$ vortices
in a stable configuration and not create vortices nearby. The latter
condition is that $B<B_{c1}$, where $B_{c1}$ is the lower critical field. Thus, the magnetic field to get the necessary flux is
\begin{align}
\pi(R_N + \lambda_{\rm eff})^2 B = N \Phi_0,
\end{align}
where $\lambda_{\rm eff}$ is the effective penetration depth of a thin
SC film defined below Eq.\ (\ref{gap1}). This gives
\begin{align}
R_N \geq \sqrt{\frac{N \Phi_0}{\pi B_{c1}}} - \lambda_{\rm eff}.\label{exp2.5}
\end{align}
Taking the standard expression for the lower critical field
$B_{c1}=(\Phi_0/4\pi\lambda_{\rm eff}^2)K_0(\kappa_{\rm eff}^{-1})$, where
$\kappa_{\rm eff}=\lambda_{\rm eff}/\xi$, Eq.\ (\ref{exp2.5}) becomes 
\begin{equation}\label{exp2.7}
 R_N\geq \lambda_{\rm eff}\left(\sqrt{2N/K_0(\kappa_{\rm eff}^{-1})}-1\right),
\end{equation}
In type-II regime $\lambda_{\rm eff}> \xi$ and Eq.\
(\ref{exp2.7}) will generally imply larger hole size than  Eq.\
(\ref{exp2}).   A larger hole size would  reduce the spectral gap to some
extent but $N$ zero modes will remain robustly present. If the SC film
remains in the type-I regime then there is no additional constraint on
$R_N$ but the applied field must be kept below the thermodynamic
critical field $B_c$ of the film.

These considerations impose some practical constraints on the material composition and
thickness $d$ of the SC film. In general we want the film to be
sufficiently thin so that scanning tunneling spectroscopy of the hole
region can be performed. On the other hand we want it to be either in
type-I or weakly type-II regime such that Eq.\ (\ref{exp2.7}) does not
enlarge the hole size significantly beyond the ideal radius given by  Eq.\
(\ref{exp2}). For Pb we have $(\xi,\lambda_L)=(83,37)$nm. Taking
$d=20$nm results in $\lambda_{\rm eff}\simeq 137$nm and  Eq.\
(\ref{exp2.7}) imposes only a mild increase in the hole size compared
to the ideal, which should not adversely affect the zero modes. For Al
we have  $(\xi,\lambda_L)=(1600,16)$nm and one can go down to very
thin films and still remain in the type-I regime. 

The TI film must be
sufficiently thick so that it exhibits well developed gapless surface
states. For Bi$_2$Se$_3$ family of materials this means thickness
larger than 5 unit cells. TI films close to this critical thickness
will also be easiest to bring to the neutrality point by back gating. 

Using a hole close to the ideal size given by Eq.\ (\ref{exp2}) will also
promote the interaction strength. Intuitively it is clear that screened
Coulomb interaction between electrons will have maximum effect on the zero
modes if their wavefunctions are packed as closely together as
possible.  With this in mind one can give a crude estimate of the
expected interaction strength $J$ as follows. Starting from Eq.\ (\ref{eta3})
with $V_0=2\pi e^2 \lambda_{\rm  TF}/\epsilon$ and using Eq.\
(\ref{phi2}) it is easy to show that 
\begin{equation}\label{exp5}
J=\left({N^3\over 3!}\overline{ J_{ijkl}^2}\right)^{1\over 2}=\sqrt{N^3\over 6}{2\pi e^2\lambda_{\rm
    TF}\over \epsilon\xi^2} {12\over M_s^{3/2}},
\end{equation}
where we identified the lengthscale $\zeta$ with the SC coherence length $\xi$.
We can obtain a physically more transparent expression by introducing
the  Bohr radius $a_0=\hbar^2/m_ee^2\simeq 0.52$\AA \ and the
corresponding Rydberg energy  $E_0=e^2/2a_0\simeq 13.6$eV, 
\begin{equation}\label{exp6}
J={48\pi\over \sqrt{6}}\sqrt{N^3\over M_s^3}
\left({a_0\lambda_{\rm TF}\over \epsilon\xi^2}\right) E_0.
\end{equation}

Several remarks are in order. Eq.\ (\ref{exp6}) implies that for a
fixed hole size $R$ the coupling strength grows as $J\sim
N^{3/2}$. It is therefore advantageous to put as many flux quanta in
the hole as can be stabilized. For the `ideal' hole size $R=R_N$ given by Eq.\ (\ref{exp2})
we have $M_s=2\pi^3N$ and the dependence on $N$ drops out. The
amplitude of $J$ will then depend only on the coherence length $\xi$,
screening length $\lambda_{\rm TF}$ and dielectric constant $\epsilon$
of the system.  To get an idea about the possible size of $J$ we
assume $\lambda_{\rm TF}\approx \xi$ and $\epsilon\approx 50$,
appropriate for the surface of a TI such as Bi$_2$Se$_3$. Eq.
(\ref{exp6}) then gives $J\approx (1{\rm \AA}/\xi)17.8$meV. It is clear that
using a superconductor with a large gap and short coherence length
would aid the observation of the SYK physics in this system at reasonable
energy and temperature scales. Taking  Pb as
a concrete example we have  $\xi\simeq 52$nm, for $d>20$nm Eq. \eqref{exp2.7} does not impose additional restrictions on the hole diameter, and one obtains $J$ in
the range of tens of $\mu$eV. This energy scale is accessible to
scanning tunneling spectroscopy (STS) which, as we argue below,
constitutes the most convenient experimental probe. 
}

\subsection{Experimental detection}
In our  proposed setup the  experimental detection of the signatures of the SYK state
can be achieved using tunneling spectroscopy. Either
 planar tunneling measurement with a fixed probe weakly coupled to the
TI surface or a scanning tunnel probe can be used. STS has the
advantage of  simultaneously
being able to image the topography of the device with
nanoscale resolution and  measure the tunneling conductance
$g(\omega)$ which is proportional to the spectral function of the
system  $A(\omega)$. A recently developed technique \cite{Ge2016}
combines an STS tip with a miniature Hall probe which allows additional
measurement of the local magnetic field $B$ at the sample surface. Such a probe is
ideally suited for the proposed SYK model setup as it can be used to
independently determine the magnetic flux and thus the number $N$ of Majorana
fermions in the system. 

 In the large-$N$ limit of the SYK model $A(\omega)$ exhibits the
characteristic $1/\sqrt{|\omega|}$ singularity (illustrated in Fig.\
\ref{fig2}a) which should be easy to distinguish from the semicircle
distribution that prevails in a system dominated by random two-fermion
tunneling terms. In the large-$N$ limit and at sufficiently low
temperature $k_BT\ll J$ the detection of the SYK behavior via
tunneling spectroscopy should therefore be relatively straightforward.

\begin{figure}[t]
\includegraphics[width = 8.5cm]{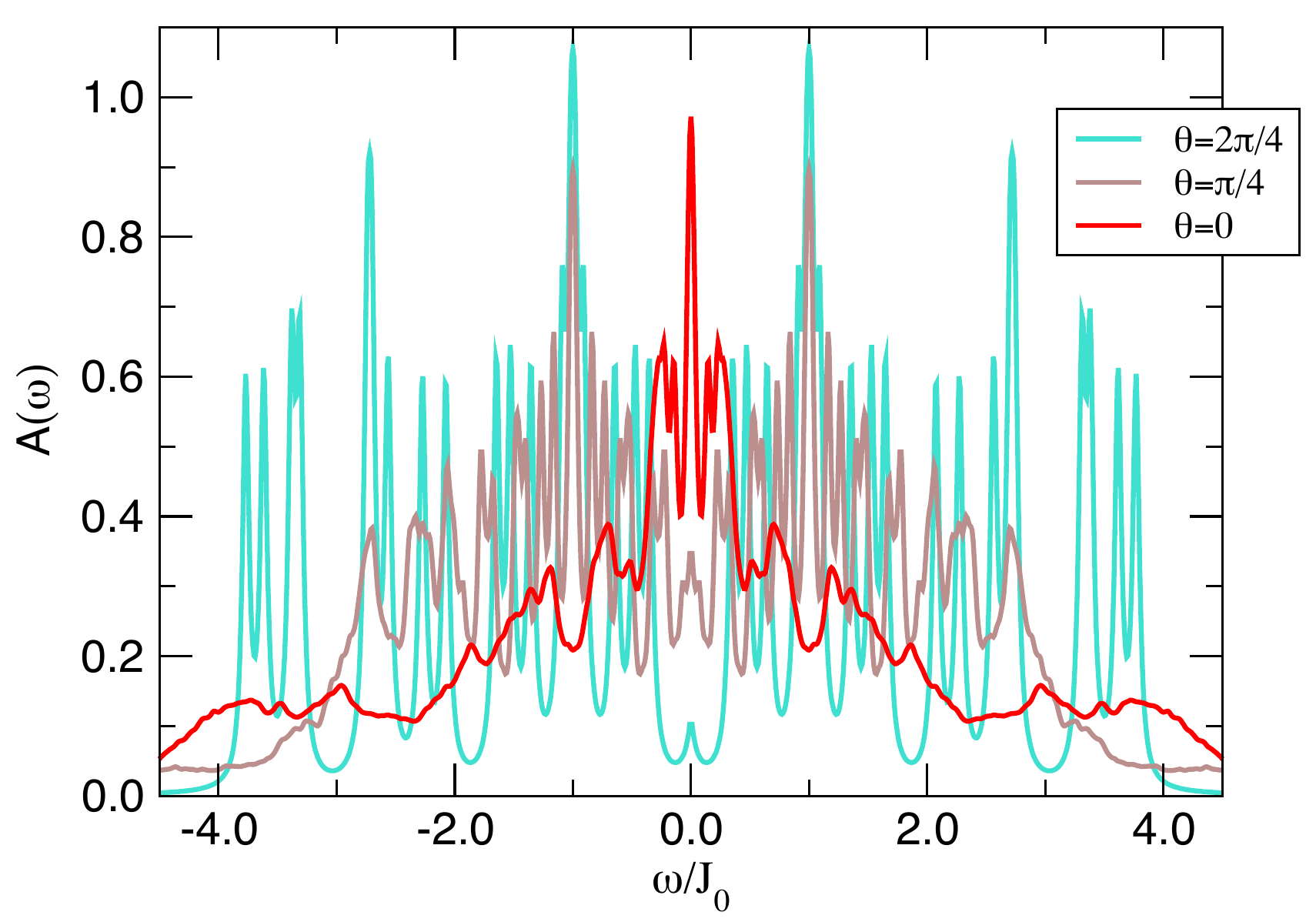}
\caption{Spectral function $A(\omega)$ calculated for the giant vortex
  geometry with $N=28$ and coupling constants 
  $(J,K)=J_0(\cos{\theta},\sin{\theta})$ taken to  interpolate
  between the fully  interacting and non-interacting limits.  
}\label{fig10}
\end{figure}
In a realistic setup the number of flux quanta $N$ will be finite and
perhaps not too large. In this case our results in Fig.\ \ref{fig7} show that the
characteristic $1/\sqrt{|\omega|}$ singularity is cut off such that $A(0)$
is finite and grows with $N$ only very slowly. Additional results
assembled in Fig. \ref{fig10} indicate that even in
this situation it is possible to distinguish the interaction-dominated
SYK behavior from the behavior characteristic of the weakly
interacting system with random two-fermion couplings. For $J\gtrsim
K$ we observe a relatively smooth spectral density peaked at $\omega=0$
characteristic of the strongly interacting regime. In the opposite limit
$J\lesssim K$ non-universal fluctuations that strongly depend on the
specific disorder realization become increasingly
prominent. Eventually, when $J\ll K$, the spectral function consists of
a series on $N$ sharp peaks. These peaks occur at the eigenvalues
of the $N\times N$ random hermitian matrix $iK_{ij}$ and represent the
single-particle excitations of the non-interacting problem at $J=0$.
At large $N$ these peaks merge to form a continuous distribution
described by the semicircle law.

Full numerical diagonalization of the SYK Hamiltonian is feasible for
$N$ up to 32 on a  desktop computer and  involves a matrix of 
size $2^{15}\times 2^{15}$ in each parity sector. By going to a supercomputer one can
plausibly reach $N=42$, \cite{boixo2016characterizing} but larger system sizes are out of reach due to
the exponential growth of the Hamiltonian matrix with $N$. Experimental
realization using the setup proposed here has no such
limitation. Measurement of the spectral function in such a system
could therefore help elucidate the approach to the large $N$ limit
in which the SYK model becomes analytically tractable by field theory
techniques. This has relevance to the spontaneous breaking of the
emergent conformal symmetry at large $N$ and a host of other interesting
issues extensively discussed in the recent literature \cite{Kitaev2015,Maldacena2016,Sachdev2015,Maldacena2016b,Hosur2016,Polchinski2016,Verbaar2016,Xu2016}.  Measurement of the out-of-time-oder correlator $F(t)$ for $N$
larger than 32 could
furthermore shed light on the emergence of the quantum chaotic
behavior in the system, scrambling and the dual relation to the extremal black hole
in AdS$_2$. A protocol to measure $F(t)$ in a system of this type is
currently unknown and this represents an interesting challenge and
an opportunity for future study.

\section{Conclusions}
To conclude, we proposed a physical realization of the
Sachdev-Ye-Kitaev model that utilizes available materials and
experimental techniques. The
proposal is to use the surface of a 3D TI at its global neutrality point
proximitized by a conventional superconductor with an irregular-shaped
hole and magnetic flux threaded through the hole. We demonstrated  that the
conventional screened Coulomb interaction between electrons in such a
setup leads to a Majorana fermion Hamiltonian at low energies with
requisite random four-fermion  couplings. Detailed analysis indicates 
behavior consistent with that expected of the SYK model. We gave
estimates for model  parameters in the realistic systems and
suggested experimental tests for the SYK behavior. This work thus provides
connections between seemingly unrelated areas of research -- mesoscopic
physics, spin liquids, general relativity, and quantum chaos -- and could lead to
experimental insights into phemomena that are of great current
interest. 

\begin{acknowledgments}

The authors are indebted to J. Alicea, O. Can, A. Kitaev, E. Lantagne-Hurtubise, M. Rozali, C.-M. Jian, I. Martin, and
S. Sachdev for illuminating discussions. We thank NSERC, CIfAR and
Max Planck - UBC Centre for Quantum Materials for support. DIP is grateful to KITP, where part of the research was conducted with support of the National Science Foundation Grant No. NSF PHY11-25915. Numerical
simulations were performed in part with computer resources provided by WestGrid and Compute Canada Calcul.
\end{acknowledgments}

\end{document}